\documentclass[preprint,12pt]{elsarticle}
\journal{Computational Materials Science}
\usepackage[utf8]{inputenc}
\usepackage[T1]{fontenc}
\biboptions{numbers,sort&compress}
\usepackage{amsmath,amssymb,amsthm,bm}
\usepackage{array,booktabs,graphicx,xcolor,tikz,placeins}
\usepackage{microtype,capt-of}
\usetikzlibrary{arrows.meta,positioning,fit,backgrounds}
\usepackage[hidelinks]{hyperref}
\definecolor{gmlBlue}{HTML}{0072B2}
\definecolor{gmlOrange}{HTML}{E69F00}
\definecolor{gmlTeal}{HTML}{009E73}

\newcommand{\E}{\mathbb E}
\newcommand{\R}{\mathbb R}
\newcommand{\dd}{\,\mathrm d}
\newcommand{\Id}{\bm I}
\newcommand{\bs}[1]{\bm{#1}}
\newcommand{\norm}[1]{\left\lVert#1\right\rVert}


\allowdisplaybreaks[1]

\newcommand{\gmlFigureWidth}{\textwidth}

\begin{document}
\begin{frontmatter}
\title{Generative marginal learning for finite-size molecular simulation with application to stress calculation}
\author[polyu-am]{Zhao Ding}
\ead{zhao.ding@polyu.edu.hk}
\author[whu-ai]{Cheng Yuan\corref{cor1}}
\ead{yuancheng@whu.edu.cn}
\cortext[cor1]{Corresponding author.}
\affiliation[polyu-am]{organization={Department of Applied Mathematics, Hong Kong Polytechnic University},
  country={Hong~Kong~SAR}}
\affiliation[whu-ai]{organization={School of Artificial Intelligence, Wuhan University},
  city={Wuhan}, postcode={430072}, state={Hubei}, country={China}}
\let\samAffiliations\elsaddress
\renewcommand{\elsaddress}{{\samAffiliations}}
\date{12 October 2026}
\begin{abstract}
We propose generative marginal learning (GML), a framework for equilibrium sampling and local observable estimation in molecular simulations. By training a conditional diffusion model on large-system simulations, GML learns the marginal distribution of a retained region, enabling configuration generation without explicitly simulating the surrounding environment or imposing periodic boundary conditions on the retained region. We derive a marginal effective potential and the associated conditional mean-force identity, which provide the basis for both denoising and force-alignment training. As an application, we estimate the spatially averaged Irving--Kirkwood stress near a vacancy defect. Numerical tests on a two-dimensional Lennard--Jones crystal and three-dimensional BCC Fe under combined tension and shear demonstrate improved local stress accuracy relative to small periodic molecular dynamics simulations.
\end{abstract}
\end{frontmatter}

\setlength{\abovedisplayskip}{8pt plus 2pt minus 2pt}
\setlength{\belowdisplayskip}{8pt plus 2pt minus 2pt}
\setlength{\abovedisplayshortskip}{3pt plus 1pt}
\setlength{\belowdisplayshortskip}{5pt plus 1pt minus 1pt}

\section{Introduction}
\label{sec:introduction}
Molecular dynamics (MD) is widely used to study material behavior, including crack growth and plastic deformation~\cite{frenkel2002,abraham1994,schiotz1998}.
Its computational cost, however, limits the system sizes accessible to direct simulation, motivating atomistic--continuum methods that retain atomic resolution near defects while describing the surrounding material at the continuum scale.
Examples include the quasicontinuum (QC) method, its finite-temperature extensions, and the heterogeneous multiscale method (HMM)~\cite{tadmor1996,dupuy2005,e2003}.

Deep learning complements this spatial reduction by reducing the cost of evaluating atomic interactions.
Neural-network potentials learn energies and forces from reference data~\cite{behler2007,zhang2018}, with recent models incorporating rotational and translational symmetries~\cite{batzner2022,batatia2022} and long-range interactions~\cite{ji2025}.
To accelerate time integration, autoregressive equivariant networks predict positions and velocities directly~\cite{thiemann2026}, while Langevin flow maps learn stochastic updates over larger time intervals~\cite{mccallum2026}.

Despite these advances, accurate sampling within a local region still requires a faithful representation of the surrounding material. This problem is particularly important near defects, where interactions with periodic images can bias local observables~\cite{cai2003}.
Absorbing boundary conditions suppress artificial reflections of lattice vibrations~\cite{li2006,yang2006,yuan2019}, while finite-temperature treatments also account for the surrounding heat bath~\cite{li2007}.
For equilibrium constitutive calculations, on the other hand, the corresponding statistical objective is to recover the local configuration distribution needed to evaluate averages of quantities such as stress.

Recent generative models offer a direct approach to sampling such distributions.
Boltzmann generators learn mappings from simple reference distributions to molecular configurations~\cite{noe2019}, with applications including crystals containing vacancies~\cite{ahmad2022}.
Diffusion models further support rare-state sampling and free-energy estimation through biasing and reweighting~\cite{xie2026}.
Related coarse-grained models learn effective distributions and forces in reduced coordinates~\cite{kohler2023}; the \emph{Two for One} framework, for example, uses diffusion models for both tasks~\cite{arts2023}.
A complementary approach is to learn marginal probabilities directly, as in generative marginalization models for discrete variables~\cite{liu2024marginalization}.
Here, we target a continuous marginal distribution that retains all atomic coordinates within a local region and represents the statistical influence of its surroundings. Unlike \emph{Two for One}, our method preserves all atomic coordinates within the retained region. The exterior coordinates are integrated out, and their statistical influence is learned without an explicit coarse-grained exterior model.

To this end, we propose generative marginal learning (GML), using a conditioned diffusion model to learn the marginal distribution of the retained region from large-system equilibrium samples. We apply the framework to estimate the spatially averaged Irving--Kirkwood (IK) stress~\cite{irving1950,admal2010}.

\begin{samepage}
\setlength{\topsep}{4pt}
\setlength{\partopsep}{0pt}
Our two main contributions are:
\begin{enumerate}
\setlength{\itemsep}{2pt}
\setlength{\parskip}{0pt}
\item We develop a generative model for a retained atomic region that captures the statistical influence of its surroundings without imposing periodic boundary conditions on the region itself. A marginal effective potential is derived to support denoising and force-alignment training.
\item As an application, we develop an estimator of the mean IK stress in an interior observation core using generated configurations and an interaction buffer. Numerical tests near vacancy clusters in two-dimensional LJ and three-dimensional BCC Fe crystals demonstrate reduced bias relative to MD simulations in small periodic cells, suggesting potential applications to multiscale constitutive calculations.
\end{enumerate}
\end{samepage}

The remainder of this paper is organized as follows.
Section~\ref{sec:marginal} formulates the marginal problem, and Section~\ref{sec:learning} introduces GML.
Section~\ref{sec:stress} develops the stress estimator, followed by numerical experiments in Section~\ref{sec:experiments}.
Section~\ref{sec:conclusions} then concludes the paper.

\section{Problem definition}
\label{sec:marginal}

Consider a large crystalline system at equilibrium, with an interior region of interest around defects such as dislocations and cracks.
We aim to calculate macroscopic physical quantities, such as stress or strain, from the atomistic information in this interior region.
To account for the influence of exterior atoms at finite temperature, one can directly simulate the full system and retain the interior degrees of freedom for thermodynamic averaging.
This approach, however, incurs a high computational cost for ensemble sampling, especially when such calculations must be repeated under different macroscopic constraints.
Alternatively, one could simulate only the small interior system with artificial boundary conditions, such as periodic boundary conditions (PBCs).
The resulting finite-size effects may reduce sampling accuracy and lead to inaccurate estimates of physical quantities.
In this work, we aim to learn an effective potential energy for the interior system that allows us to sample the same marginal distribution as full-system simulation.

\begin{figure}[!tbp]
\centering
\begin{tikzpicture}[font=\fontsize{9}{11}\selectfont,>=Latex,line width=0.7pt]
  \fill[gmlBlue!7] (0,0) rectangle (6.6,6.1);
  \fill[gmlBlue!24] (0.9,1.0) rectangle (5.7,5.1);
  \fill[gmlOrange!48] (2.0,2.1) rectangle (4.6,3.9);
  \foreach \j in {1,...,15} {
    \foreach \i in {1,...,16} {
      \pgfmathsetmacro{\xx}{0.10+0.38*\i+0.19*mod(\j,2)}
      \pgfmathsetmacro{\yy}{0.08+0.38*\j}
      \fill[gmlBlue!65] (\xx,\yy) circle (0.025);
    }
  }
  \draw[gmlBlue!70!black,thick] (0,0) rectangle (6.6,6.1);
  \draw[gmlBlue,thick] (0.9,1.0) rectangle (5.7,5.1);
  \draw[gmlOrange!85!black,thick] (2.0,2.1) rectangle (4.6,3.9);
  \node[font=\fontsize{10}{12}\selectfont,inner sep=0pt] at (3.33,5.59) {$\Omega$};
  \node[font=\fontsize{10}{12}\selectfont,inner sep=0pt] at (3.33,4.45) {$R$};
  \node[font=\fontsize{10}{12}\selectfont,inner sep=0pt] at (3.33,2.93) {$C$};
  \node[anchor=west,align=left] at (7.1,5.55)
    {\textbf{Full system $\Omega$}\\$N$ atoms; coordinates $(x,y)$};
  \node[anchor=west,align=left] at (7.1,4.30)
    {\textbf{Retained region $R$}\\$M<N$ atoms; coordinates $x$\\Learn their joint distribution};
  \node[anchor=west,align=left] at (7.1,3.05)
    {\textbf{Observation core $C$}\\Evaluate physical quantities here};
  \node[anchor=west,align=left] at (7.1,1.95)
    {\textbf{Interaction buffer $R\setminus C$}\\Supplies interacting neighbors};
  \node[anchor=west,align=left] at (7.1,0.70)
    {\textbf{Exterior $\Omega\setminus R$}\\Integrate out coordinates $y$\\Generate only the retained region};
\end{tikzpicture}
\caption{The domain reduction used by GML.
The full domain $\Omega$ contains $N$ atoms.
The retained region $R\subset\Omega$ contains $M$ atoms with coordinates $x$; the exterior coordinates $y$ are integrated out.
The observation core $C\subset R$ is surrounded by the interaction buffer $R\setminus C$.
GML learns the joint distribution of all retained coordinates from full-system samples, while observables are evaluated in $C$.
The coordinates shown here precede the translation alignment used for training (Fig.~\ref{fig:learning-flow}).
Atomic dots and region sizes are schematic.}
\label{fig:setting}
\end{figure}
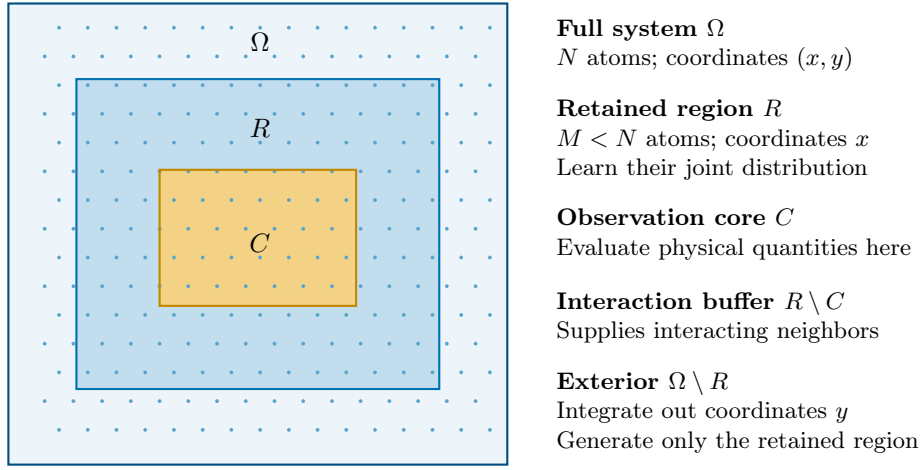

To begin with, let $\Omega$ denote the full spatial domain and $C$ the observation core where these quantities are evaluated (see Fig.~\ref{fig:setting}).
The retained region $R$ satisfies $C\subset R\subset\Omega$.
The layer $R\setminus C$ serves as an interaction buffer that supplies the neighbors needed to evaluate observables near the core boundary.
Let $N$ and $M$ be the numbers of full-system and retained atoms, respectively, with typically $M\ll N$.
Denote the fixed index set of the retained atoms by $\mathcal I_R=\{1,\ldots,M\}$.
The retained coordinate vector is $x=(q_1,\ldots,q_M)\in\R^{dM}$, and the exterior coordinate vector is $y=(q_{M+1},\ldots,q_N)\in\R^{d(N-M)}$, where $d$ is the spatial dimension and $q_i\in\R^d$ is the position of atom $i$.
At finite temperature, the full configuration $q=(x,y)\in\R^{dN}$ in $\Omega$ can be expressed in terms of a homogeneous deformation and atomic displacements~\cite{dai2022}:
\begin{equation}
 q_i=A X_i+u_i, \quad i=1,\ldots,N,
 \label{eq:displacement}
\end{equation}
where $X_i\in\R^d$ denotes the fixed reference position, $u_i\in\R^d$ is the displacement, and $A\in\R^{d\times d}$ is a homogeneous deformation matrix.

For the canonical ensemble at fixed particle number $N$, volume, and temperature $T$ (NVT), the full-system configurational density is
\begin{equation}
 p(x,y\mid A,T)=\widetilde Z^{-1}
                 e^{-\beta\widetilde V(x,y;A)},\qquad
 \widetilde Z=\int e^{-\beta\widetilde V(x,y;A)}\dd x\dd y.
 \label{eq:full-density}
\end{equation}
Here $\beta=(k_B T)^{-1}$, where $k_B$ is Boltzmann's constant. The function $\widetilde V(x,y;A)$ is the potential energy of the full system, which is typically subject to periodic or fixed boundary conditions.
We also write $p_\Omega(q\mid A,T)$ for this full-system equilibrium density in terms of the full coordinate vector $q$.

To obtain the exact marginal distribution of the interior coordinates $x$, we integrate out the exterior degrees of freedom, leading to the following partition function, effective potential energy, and marginal density:

\begin{align}
 Z_y(x;A,T)&=\int e^{-\beta\widetilde V(x,y;A)}\dd y,\label{eq:conditional-partition}\\
 V(x;A,T)&=-k_B T\log Z_y(x;A,T),\label{eq:effective-potential}\\
 p_R(x\mid A,T)&=\int p(x,y\mid A,T)\dd y
                =Z^{-1}e^{-\beta V(x;A,T)},\label{eq:marginal-density}\\
 Z&=\int e^{-\beta V(x;A,T)}\dd x=\widetilde Z.
 \label{eq:partition-equality}
\end{align}
Here $V$ is also called a potential of mean force: it accounts for the influence of the exterior after its coordinates have been integrated out~\cite{noid2008}. It generally depends on both the temperature and the full-system boundary conditions.

For any integrable observable $\psi(x)$ of the retained coordinates, ensemble averaging at fixed $A,T$ gives
\begin{equation}
 \begin{aligned}
 \langle \psi\rangle_{\Omega}(A,T)
 &:=\int \psi(x)p(x,y\mid A,T)\dd x\dd y\\
 &=\int \psi(x)p_R(x\mid A,T)\dd x
 =:\langle \psi\rangle_R(A,T).
 \end{aligned}
 \label{eq:observable-marginal}
\end{equation}
The averages depend on the prescribed deformation $A$ and temperature $T$.
In the following, $\E_\Omega$ denotes expectation over $(x,y)\sim p(\cdot,\cdot\mid A,T)$, and $\E_R$ denotes expectation over $x\sim p_R(\cdot\mid A,T)$.

According to Eq.~\eqref{eq:observable-marginal}, once we know $p_R(\cdot\mid A,T)$ or can draw samples from it, we can calculate interior statistical averages without further full-system simulations.

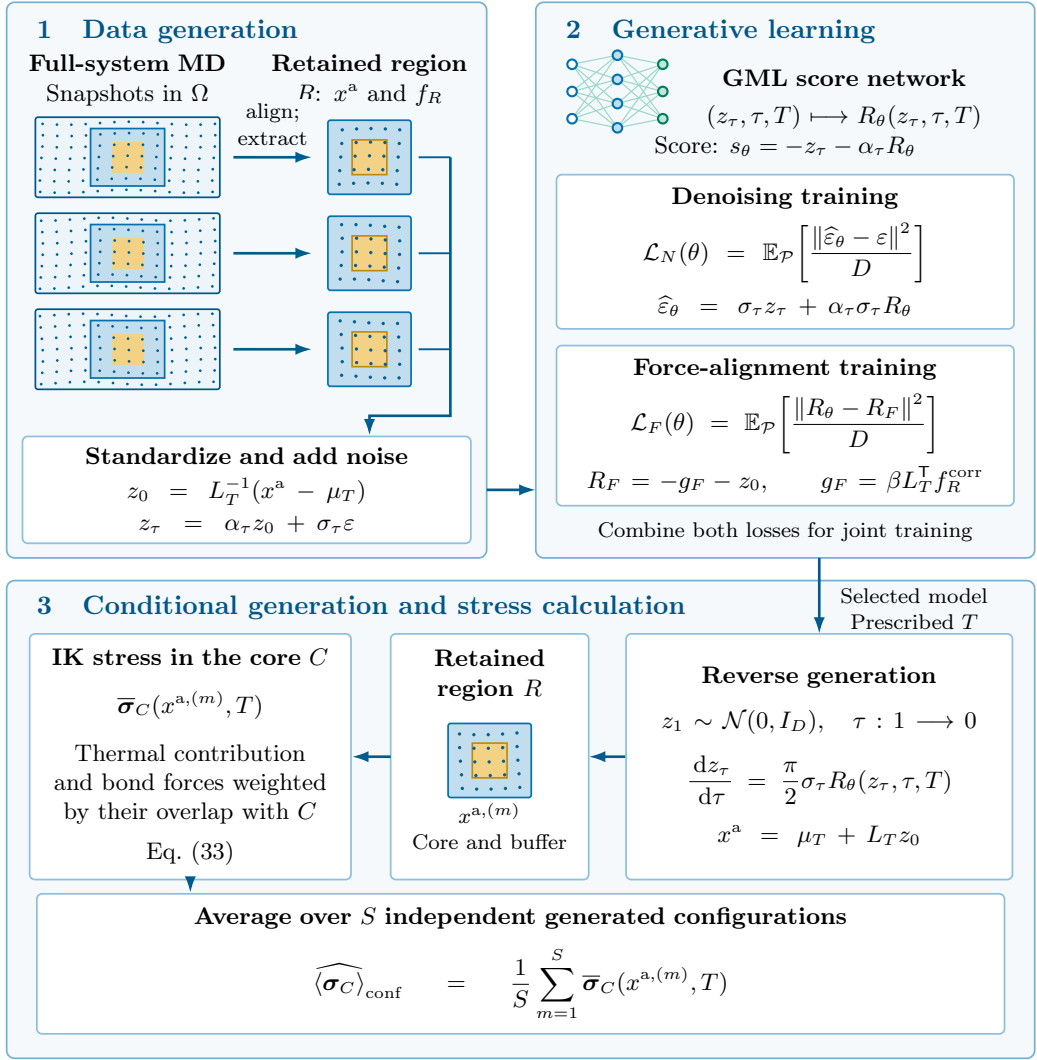
\begin{figure}[!tbp]
\centering
\resizebox{0.96\textwidth}{!}{\begingroup
\linespread{1}\selectfont
\newcommand{\gmlFullSnapshot}[3]{%
  \begin{scope}[shift={(#1,#2)}]
    \filldraw[fill=gmlBlue!7,draw=gmlBlue!75!black,rounded corners=1pt]
      (-1.22,-0.53) rectangle (1.22,0.53);
    \filldraw[fill=gmlBlue!24,draw=gmlBlue,line width=0.65pt]
      (-0.49,-0.38) rectangle (0.49,0.38);
    \fill[gmlOrange!48] (-0.21,-0.20) rectangle (0.21,0.20);
    \foreach \i in {-6,...,6}{\foreach \j in {-3,...,3}{
        \fill[gmlBlue!85!black] ({0.19*\i+0.02*sin(43*\j+#3)},
          {0.15*\j+0.02*cos(41*\i+#3)}) circle (0.018);
    }}
  \end{scope}}
\newcommand{\gmlRetainedSnapshot}[3]{%
  \begin{scope}[shift={(#1,#2)}]
    \filldraw[fill=gmlBlue!24,draw=gmlBlue,rounded corners=1pt]
      (-0.55,-0.49) rectangle (0.55,0.49);
    \filldraw[fill=gmlOrange!48,draw=gmlOrange!85!black,line width=0.6pt]
      (-0.23,-0.23) rectangle (0.23,0.23);
    \foreach \i in {-2,...,2}{\foreach \j in {-2,...,2}{
        \fill[gmlBlue!85!black] ({0.19*\i+0.02*sin(43*\j+#3)},
          {0.18*\j+0.02*cos(41*\i+#3)}) circle (0.022);
    }}
  \end{scope}}
\begin{tikzpicture}[
  font=\fontsize{9}{11}\selectfont,
  line width=0.65pt,
  >=Latex,
  flow/.style={-{Latex[length=2.2mm,width=1.5mm]},line width=1pt,
    draw=gmlBlue!80!black},
  panel/.style={rounded corners=3pt,draw=gmlBlue!55,fill=gmlBlue!4},
  card/.style={rounded corners=2pt,draw=gmlBlue!45,fill=white},
  small/.style={font=\fontsize{8}{10}\selectfont,align=center},
  title/.style={font=\fontsize{10}{12}\selectfont\bfseries,align=left},
  mathbox/.style={rounded corners=2pt,draw=gmlBlue!45,
    fill=white,align=center,inner sep=5pt}
]
\draw[panel] (0,5.90) rectangle (6.35,13.25);
\draw[panel] (7.00,5.90) rectangle (13.60,13.25);
\draw[panel] (0,-0.72) rectangle (13.60,5.60);
\node[title,anchor=west,text=gmlBlue!75!black] at (0.25,12.87)
  {1\quad Data generation};
\node[title,anchor=west,text=gmlBlue!75!black] at (7.25,12.87)
  {2\quad Generative learning};
\node[align=center,text width=2.65cm,inner sep=1pt] at (1.60,12.22) {\textbf{Full-system MD}\\Snapshots in $\Omega$};
\node[align=center,text width=2.65cm,inner sep=1pt] at (4.80,12.22) {\textbf{Retained region}\\$R$: $x^{\mathrm a}$ and $f_R$};

\gmlFullSnapshot{1.60}{11.20}{0}
\gmlFullSnapshot{1.60}{9.93}{60}
\gmlFullSnapshot{1.60}{8.66}{120}
\gmlRetainedSnapshot{4.80}{11.20}{0}
\gmlRetainedSnapshot{4.80}{9.93}{60}
\gmlRetainedSnapshot{4.80}{8.66}{120}
\foreach \yy in {11.20,9.93,8.66}{\draw[flow] (3.00,\yy)--(4.04,\yy);}
\node[small,text width=1.03cm,fill=gmlBlue!4,inner sep=1.5pt]
  at (3.52,11.64) {align;\\extract};

\node[card,
  align=center,text width=5.65cm,minimum height=1.18cm,inner sep=4pt]
  (prepare) at (3.175,6.80) {
  \textbf{Standardize and add noise}\\[2pt]
  $z_0=L_T^{-1}(x^{\mathrm a}-\mu_T)$\\[1pt]
  $z_\tau=\alpha_\tau z_0+\sigma_\tau\varepsilon$};
\foreach \yy in {11.20,9.93,8.66}{
  \draw[draw=gmlBlue!80!black,line width=0.7pt] (5.45,\yy)--(5.87,\yy);
}
\coordinate (collected) at (4.80,7.78);
\draw[flow] (5.87,11.20)--(5.87,7.78)--(collected)--(collected |- prepare.north);
\draw[flow] (6.35,6.80)--(7.00,6.80);

\begin{scope}[shift={(7.48,12.07)}]
  \foreach \a in {-0.36,0,0.36}{\foreach \b in {-0.48,-0.16,0.16,0.48}{
    \draw[gmlTeal!40,line width=0.45pt] (0,\a)--(0.60,\b);
    \draw[gmlTeal!40,line width=0.45pt] (0.60,\b)--(1.20,\a);
  }}
  \foreach \a in {-0.36,0,0.36}{
    \filldraw[fill=white,draw=gmlBlue,line width=0.8pt] (0,\a) circle (0.063);
    \filldraw[fill=gmlTeal!25,draw=gmlTeal!80!black,line width=0.8pt]
      (1.20,\a) circle (0.063);
  }
  \foreach \b in {-0.48,-0.16,0.16,0.48}{
    \filldraw[fill=gmlBlue!24,draw=gmlBlue,line width=0.8pt]
      (0.60,\b) circle (0.063);
  }
\end{scope}
\node[align=center,text width=3.9cm] at (11.08,11.97)
  {\textbf{GML score network}\\[3pt]
   $(z_\tau,\tau,T)\longmapsto R_\theta(z_\tau,\tau,T)$};
\node[align=center] at (10.30,11.35)
  {Score: $s_\theta=-z_\tau-\alpha_\tau R_\theta$};

\node[mathbox,text width=5.70cm,minimum height=1.48cm]
  at (10.30,9.95) {
  \textbf{Denoising training}\\[4pt]
  $\displaystyle\mathcal L_N(\theta)
    =\E_{\mathcal P}\!\left[\frac{\norm{\widehat\varepsilon_\theta-\varepsilon}^2}{D}\right]$\\[4pt]
  $\widehat\varepsilon_\theta=\sigma_\tau z_\tau+
    \alpha_\tau\sigma_\tau R_\theta$};
\node[mathbox,text width=5.70cm,minimum height=1.72cm]
  at (10.30,7.65) {
  \textbf{Force-alignment training}\\[4pt]
  $\displaystyle\mathcal L_F(\theta)
    =\E_{\mathcal P}\!\left[\frac{\norm{R_\theta-R_F}^2}{D}\right]$\\[4pt]
  $R_F=-g_F-z_0,\qquad g_F=\beta L_T^{\mathsf T}f_R^{\rm corr}$};
\node[small] at (10.30,6.25) {Combine both losses for joint training};

\node[title,anchor=west,text=gmlBlue!75!black] at (0.25,5.25)
  {3\quad Conditional generation and stress calculation};
\draw[flow] (10.75,5.90)--(10.75,4.895);
\node[small,fill=gmlBlue!4,inner sep=2pt,anchor=west] at (10.95,5.23)
  {Selected model\\Prescribed $T$};

\begin{scope}[yshift=-0.25cm]
\node[card,align=center,text width=4.76cm,
  minimum height=3.25cm,inner sep=5pt] (reverse) at (10.75,3.52) {
  \textbf{Reverse generation}\\[6pt]
  $z_1\sim\mathcal N(0,I_D),\quad\tau:1\longrightarrow0$\\[7pt]
  $\displaystyle\frac{\dd z_\tau}{\dd\tau}
     =\frac{\pi}{2}\sigma_\tau R_\theta(z_\tau,\tau,T)$\\[7pt]
  $x^{\mathrm a}=\mu_T+L_Tz_0$};

\draw[card] (5.08,1.895) rectangle (7.70,5.145);
\node[align=center] at (6.39,4.60) {\textbf{Retained}\\\textbf{region $R$}};
\gmlRetainedSnapshot{6.39}{3.47}{200}
\node[small] at (6.39,2.60) {$x^{\mathrm a,(m)}$\\Core and buffer};
\draw[flow] (reverse.west)--(7.70,3.52);

\node[card,
  align=center,text width=3.90cm,minimum height=3.25cm,inner sep=5pt]
  (stress) at (2.43,3.52) {
  \textbf{IK stress in the core $C$}\\[7pt]
  $\overline{\bs\sigma}_C(x^{\mathrm a,(m)},T)$\\[7pt]
  Thermal contribution\\and bond forces weighted\\by their overlap with $C$\\[4pt]
  Eq.~\eqref{eq:ik-config}};
\draw[flow] (5.08,3.52)--(stress.east);

\node[card,
  align=center,text width=12.45cm,minimum height=1.43cm,inner sep=5pt]
  (average) at (6.80,0.78) {
  \textbf{Average over $S$ independent generated configurations}\\[6pt]
  $\displaystyle\widehat{\langle\bs\sigma_C\rangle}_{\rm conf}
    =\frac{1}{S}\sum_{m=1}^{S}\overline{\bs\sigma}_C(x^{\mathrm a,(m)},T)$};
\draw[flow] (stress.south)--(stress.south |- average.north);
\end{scope}
\end{tikzpicture}%
\endgroup%
}
\caption{Data generation, training, and stress estimation with GML.
Blue and orange denote $R$ and $C$ as in Fig.~\ref{fig:setting}; icons are schematic.
Full-system molecular dynamics supplies configurations and, for force-alignment training, forces $f_R$ including exterior interactions.
Translation alignment gives the retained coordinates $x^{\mathrm a}$ defined in Eq.~\eqref{eq:aligned-coordinates}; corrected forces follow Eq.~\eqref{eq:corrected-force}.
Preprocessing follows Eqs.~\eqref{eq:standardization} and~\eqref{eq:forward}; network and loss notation follows Section~\ref{sec:training}.
The losses average over temperature, full-system configuration, noise level, and Gaussian noise according to $\mathcal P$ in Eq.~\eqref{eq:training-law}, with $D=dM$.
At fixed $T$, $S$ independent samples $x^{\mathrm a,(m)}$, $m=1,\ldots,S$, yield the buffered mean IK stress via Eq.~\eqref{eq:configuration-route}.}
\label{fig:learning-flow}
\end{figure}

\section{Generative marginal learning}
\label{sec:learning}
In this work, we employ a diffusion model to learn the conditional marginal distribution $p_R(\cdot\mid A,T)$, following the workflow in Fig.~\ref{fig:learning-flow}. To this end, we use the score function of the target distribution, defined as the gradient of its log-density with respect to the retained coordinates:
\begin{equation}
 s_R(x;A,T)=\nabla_x\log p_R(x\mid A,T).
 \label{eq:clean-score}
\end{equation}
It describes how the density changes under a small change of the retained coordinates.
The corresponding effective force is defined by
\begin{equation}
  f^{\rm eff}(x;A,T)=-\nabla_x V(x;A,T).
\end{equation}
Differentiating Eq.~\eqref{eq:effective-potential}, assuming differentiation under the exterior integral is valid, gives
\nopagebreak[4]
\begin{align}
 \nabla_x V(x;A,T)
 &=-k_B T\,Z_y(x;A,T)^{-1}\nabla_x Z_y(x;A,T)\nonumber\\
 &=Z_y(x;A,T)^{-1}\int \nabla_x\widetilde V(x,y;A)
                  e^{-\beta\widetilde V(x,y;A)}\dd y\nonumber\\
 &=\E_\Omega[\nabla_x\widetilde V(x,y;A)\mid x].
 \label{eq:mean-force-proof}
\end{align}
It follows that
\begin{equation}
 \begin{aligned}
 f^{\rm eff}(x;A,T)&=\E_\Omega[f_R(x,y;A)\mid x],\\
 s_R(x;A,T)&=\beta f^{\rm eff}(x;A,T),
 \end{aligned}
 \label{eq:mean-force}
\end{equation}
where $f_R(x,y;A):=-\nabla_x\widetilde V(x,y;A)$ is the force on the retained atoms in the full system.
This force includes exterior interactions; the conditional expectation averages over $y$ at fixed $x,A,T$.

This identity supports force-alignment training as an alternative to denoising training for the conditional diffusion model.

\noindent\textbf{Remark (Ensembles and conditioning).}
The framework extends to other equilibrium ensembles by using their macroscopic constraints as conditioning variables $c$ and learning the score $s_R(x;c)=\nabla_x\log p_R(x\mid c)$.
For the isothermal--isobaric (NPT) ensemble, the constraints are particle number $N$, imposed pressure $P$, and temperature $T$, so $c=(N,P,T)$.
For simplicity, this work fixes $N$ and $A$ and considers conditional generation at different temperatures $T$.
Below, we suppress the fixed $N$ and $A$ in diffusion and network notation, while keeping $T$ explicit.
In particular, the standardized density $p_0(z\mid T)$ introduced in Eq.~\eqref{eq:standard-density} is shorthand for $p_0(z\mid N, A,T)$ at the prescribed $N$, $A$.

\subsection{Data generation and forward process}
Before training, we perform molecular dynamics simulations of the full system $\Omega$ with periodic boundary conditions.
After equilibration, we extract configurations of the retained region $R$ from the trajectories and use them as training samples for the conditional generative model.
In these simulations, the crystal can drift as a whole because its potential energy is invariant under uniform translations~\cite{ahmad2022}.
Using consistently unwrapped coordinates, we subtract the full-system mean displacement:
\begin{equation}
 \bar u=\frac1N\sum_{j=1}^N(q_j-A X_j),\qquad
 x_i^{\mathrm a}=q_i-\bar u,\quad i\in\mathcal I_R,
 \label{eq:aligned-coordinates}
\end{equation}
where $x^{\mathrm a}=(x_i^{\mathrm a})_{i\in\mathcal I_R}$ contains the aligned coordinates. Removing the common translation lets the model focus on thermal fluctuations relative to the lattice.
Let $p_R^{\mathrm a}(\cdot\mid A,T)$ denote the marginal density of the retained coordinates after the full configuration has been aligned.
Its relation to the original full-system density is
\begin{align*}
 p_R^{\mathrm a}(x^{\mathrm a}\mid A,T)
 &=\int p_\Omega(q\mid A,T)\,
       \delta_{dM}\bigl(x^{\mathrm a}-x^{\mathrm a}(q)\bigr)\dd q\\
 &=\frac{1}{\widetilde Z}\int e^{-\beta\widetilde V(q;A)}\,
       \delta_{dM}\bigl(x^{\mathrm a}-x^{\mathrm a}(q)\bigr)\dd q.
\end{align*}
Here $x^{\mathrm a}(q)$ is the aligned retained vector defined in Eq.~\eqref{eq:aligned-coordinates}, and $\delta_{dM}$ is the $dM$-dimensional Dirac delta.
The delta fixes the aligned retained coordinates, while the integral removes the remaining degrees of freedom, including the common translation.
The corresponding potential of mean force is $V^{\mathrm a}(x^{\mathrm a};A,T)=-k_B T\log p_R^{\mathrm a}(x^{\mathrm a}\mid A,T)+\mathrm{const}$, where the additive constant is independent of $x^{\mathrm a}$.
Because $\bar u$ depends on both retained and exterior coordinates, $p_R^{\mathrm a}$ generally differs from the unaligned marginal $p_R$.
Accordingly, $V^{\mathrm a}$ is distinct from $V$ in Eq.~\eqref{eq:effective-potential}.

To reduce correlations and balance fluctuation scales before adding isotropic noise, we use covariance whitening, a standard preprocessing step~\cite{kessy2018} also used in molecular Boltzmann generators~\cite{noe2019}:
\begin{equation}
 \Sigma_T=L_TL_T^{\mathsf T},\qquad
 z_0=L_T^{-1}(x^{\mathrm a}-\mu_T)\in\R^D,\quad D=dM.
 \label{eq:standardization}
\end{equation}
Here $\mu_T$ and $\Sigma_T$ are the empirical mean and regularized covariance of the aligned training configurations at temperature $T$.
The matrix $L_T$ is the lower-triangular Cholesky factor with positive diagonal entries.
The superscript $\mathsf T$ denotes matrix transpose.
The inverse transformation, $x^{\mathrm a}=\mu_T+L_Tz_0$, gives the standardized conditional density
\begin{equation}
 p_0(z\mid T)=\lvert\det L_T\rvert\,p_R^{\mathrm a}(\mu_T+L_Tz\mid A,T).
 \label{eq:standard-density}
\end{equation}
To sample from the density in Eq.~\eqref{eq:standard-density}, we use variance-preserving (VP) diffusion~\cite{song2021} with an unshifted continuous version of the cosine noise schedule~\cite{nichol2021} as the forward process.
With Gaussian noise $\varepsilon\sim\mathcal N(0,I_D)$ independent of $z_0$, the forward perturbation is
\begin{equation}
 z_\tau=\alpha_\tau z_0+\sigma_\tau\varepsilon,\qquad
 \alpha_\tau=\cos\frac{\pi\tau}{2},\quad
 \sigma_\tau=\sin\frac{\pi\tau}{2}.
 \label{eq:forward}
\end{equation}
Here $\tau\in[0,1]$ is the noise level, $I_D$ is the identity matrix, and $\mathcal N(0,I_D)$ is the standard Gaussian distribution.
Thus $z_0\sim p_0(\cdot\mid T)$ and $z_1\sim\mathcal N(0,I_D)$.
Training samples follow the joint law $\mathcal P$:
\begin{equation}
 T\sim\pi_{\rm tr},\quad q\sim p_\Omega(\cdot\mid A,T),\quad
 \tau\sim\mathcal U(0,1),\quad\varepsilon\sim\mathcal N(0,I_D).
 \label{eq:training-law}
\end{equation}
The temperature distribution $\pi_{\rm tr}$ is uniform over the training temperatures, and $\mathcal U(0,1)$ denotes the uniform distribution on $(0,1)$.
The variables $\tau$ and $\varepsilon$ are independent of each other and of $(T,q)$.

\subsection{Conditional generation using the reverse process}
\label{sec:reverse-generation}
Let $p_\tau(\cdot\mid T)$ denote the density of $z_\tau$ and $s_\tau(z;T)=\nabla_z\log p_\tau(z\mid T)$ its conditional score.
For the schedule in Eq.~\eqref{eq:forward}, define $\kappa_\tau=\pi\tan(\pi\tau/2)$.
At a prescribed temperature $T$, the forward and reverse processes satisfy the stochastic differential equations (SDEs)~\cite{song2021}
\begin{align}
 \text{Forward process:}\quad
 &\dd z_\tau=-\tfrac12\kappa_\tau z_\tau\dd\tau
   +\sqrt{\kappa_\tau}\dd W_\tau,\label{eq:forward-sde}\\*
 \text{Reverse process:}\quad
 &\begin{aligned}[t]
 \dd z_\tau={}&\left[-\tfrac12\kappa_\tau z_\tau
   -\kappa_\tau s_\tau(z_\tau;T)\right]\dd\tau\\
 &+\sqrt{\kappa_\tau}\dd\overline W_\tau.
 \end{aligned}
 \label{eq:reverse-sde}
\end{align}
Here $W_\tau$ and $\overline W_\tau$ are $D$-dimensional Brownian motions for the forward and reverse equations, with component increment variances $\dd\tau$ and $|\dd\tau|$, respectively.
The forward equation is defined for $0\le\tau<1$, while the reverse equation runs toward $\tau=0$.
With the exact score, solving Eq.~\eqref{eq:reverse-sde} backward from the Gaussian limit at $\tau=1$ recovers $p_0(\cdot\mid T)$.

In this work, we use the probability-flow ordinary differential equation (ODE), which requires no additional noise after initialization and can be solved with standard ODE methods.
The ODE formulation also offers a potential route to one-step generation through consistency models.
These models can be trained from an existing diffusion model to map points on each ODE trajectory directly to its data endpoint~\cite{song2023consistency}.
With the exact score, the ODE has the same time-dependent marginal densities as the reverse SDE in Eq.~\eqref{eq:reverse-sde}~\cite{song2021}:
\begin{equation}
 \frac{\dd z_\tau}{\dd\tau}
 =-\tfrac12\kappa_\tau\left[z_\tau+s_\tau(z_\tau;T)\right].
 \label{eq:probability-flow}
\end{equation}
For each independent draw $z_1\sim\mathcal N(0,I_D)$, we integrate Eq.~\eqref{eq:probability-flow} backward to $\tau=0$ and recover the aligned interior configuration $x^{\mathrm a}=\mu_T+L_Tz_0$.
With the exact score and integration, this yields samples from $p_R^{\mathrm a}(\cdot\mid A,T)$.
In practice, however, this conditional score must be learned using suitable training objectives.

\subsection{Training objectives}
\label{sec:training}

To learn the score required for generation, we use a $D$-component neural network $R_\theta$ with trainable parameters $\theta$.
The network takes the noisy configuration $z_\tau$, noise level $\tau$, and temperature $T$ as inputs and learns a residual correction:
\begin{equation}
 s_\theta(z_\tau,\tau,T)
 :=-z_\tau-\alpha_\tau R_\theta(z_\tau,\tau,T)
 \approx s_\tau(z_\tau;T).
 \label{eq:residual-score}
\end{equation}
The motivation for this construction is that the whitened data have approximately zero mean and unit covariance (Eq.~\eqref{eq:standardization}).
We therefore use the standard Gaussian as a reference and learn a correction to its known score $-z_\tau$, following the idea of combining Gaussian and neural scores~\cite{vahdat2021}.
This parameterization also simplifies the generation step.
Substituting $s_\theta$ for $s_\tau$ in Eq.~\eqref{eq:probability-flow} and using $\kappa_\tau\alpha_\tau=\pi\sigma_\tau$ gives the sampling equation
\begin{equation}
 \frac{\dd z_\tau}{\dd\tau}
 =\frac{\pi}{2}\sigma_\tau R_\theta(z_\tau,\tau,T).
 \label{eq:learned-probability-flow}
\end{equation}
For finite $R_\theta$, this drift remains finite at $\tau=1$.

To learn this residual correction, we consider denoising and force alignment.
Both losses average over the joint training law $\mathcal P$ in Eq.~\eqref{eq:training-law}.

\subsubsection{Denoising loss}

The first objective is motivated by the relationship between the conditional score and the Gaussian noise introduced in the forward process.
For $0<\tau<1$, differentiating the transition density in Eq.~\eqref{eq:forward} and averaging over clean configurations yield~\cite{vincent2011}
\begin{equation}
 s_\tau(z_\tau;T)
 =-\sigma_\tau^{-1}\E_{\mathcal P}[\varepsilon\mid z_\tau,\tau,T].
 \label{eq:denoising-score}
\end{equation}
This identity allows score estimation to be reformulated as a noise prediction problem~\cite{ho2020}:
\begin{align}
 \widehat\varepsilon_\theta
 &:=-\sigma_\tau s_\theta
 =\sigma_\tau z_\tau+\alpha_\tau\sigma_\tau R_\theta,\nonumber\\
 \mathcal L_N(\theta)
 &=\E_{\mathcal P}\left[\frac{\norm{\widehat\varepsilon_\theta-\varepsilon}^2}{D}\right].
 \label{eq:denoising-loss}
\end{align}
Here, all network outputs are evaluated at $(z_\tau,\tau,T)$, and $\norm{\cdot}$ denotes the Euclidean norm.

\subsubsection{Force-alignment loss}
When full-system forces are available, the mean-force relation in Eq.~\eqref{eq:mean-force} provides an alternative training objective.
Because the alignment in Eq.~\eqref{eq:aligned-coordinates} depends on the full configuration, we first correct the force labels for differentiation with respect to $x^{\mathrm a}$~\cite{noid2008,kalligiannaki2015}.

To obtain this correction, change one retained coordinate $x_i^{\mathrm a}$ by a small displacement $\bs\delta\in\R^d$, keeping the others fixed.
For $M<N$, shifting each exterior atom by $-\bs\delta/(N-M)$ keeps the full-system mean displacement fixed.
With physical forces $\bs f_i=-\nabla_{q_i}\widetilde V$ satisfying $\sum_{i=1}^N\bs f_i=0$, the first-order energy change is $-\bs f_i^{\rm corr}\cdot\bs\delta$, where
\begin{equation}
 \bs f_i^{\rm corr} =\bs f_i-\frac{\sum_{j\notin\mathcal I_R}\bs f_j}{N-M} =\bs f_i+\frac{\sum_{j\in\mathcal I_R}\bs f_j}{N-M},
 \qquad i\in\mathcal I_R.
 \label{eq:corrected-force}
\end{equation}
Let $f_R^{\rm corr}:=(\bs f_i^{\rm corr})_{i\in\mathcal I_R}\in\R^{dM}$.
Applying Eq.~\eqref{eq:mean-force-proof} along these displacements, with vanishing integration boundary terms, gives
\begin{equation}
 \begin{aligned}
 s_R^{\mathrm a}(x^{\mathrm a};T)
 &:=\nabla_{x^{\mathrm a}}\log p_R^{\mathrm a}(x^{\mathrm a}\mid A,T)\\
 &=-\beta\nabla_{x^{\mathrm a}}V^{\mathrm a}(x^{\mathrm a};A,T)\\
 &=-\beta\E_\Omega[\nabla_{x^{\mathrm a}}\widetilde V\mid x^{\mathrm a}]\\
 &=\beta\E_\Omega[f_R^{\rm corr}\mid x^{\mathrm a}].
 \end{aligned}
 \label{eq:centered-score}
\end{equation}
So $f_R^{\rm corr}$ provides score labels.
Under $x^{\mathrm a}=\mu_T+L_Tz$, Eq.~\eqref{eq:standard-density} gives the standardized labels and score:
\begin{equation}
 g_F:=\beta L_T^{\mathsf T}f_R^{\rm corr},\qquad g(z;T):=\nabla_z\log p_0(z\mid T)=L_T^{\mathsf T}s_R^{\mathrm a}(x^{\mathrm a};T).
 \label{eq:standard-score}
\end{equation}
Equation~\eqref{eq:centered-score} then gives $\E_\Omega[g_F\mid z_0,T]=g(z_0;T)$.
Thus, $g_F$ labels the zero-noise score $g=s_0$.
To extend this supervision to noisy inputs, let $k_\tau(z\mid z_0)$ denote the Gaussian transition density in Eq.~\eqref{eq:forward}.
For $0<\tau<1$, independent forward noise and integration by parts give~\cite{debortoli2024,ko2026}
\begingroup
\allowdisplaybreaks[4]
\begin{align*}
 &\E_{\mathcal P}[g_F\mid z_\tau=z,\tau,T]\\*
 &\quad=\E_{\mathcal P}[g(z_0;T)\mid z_\tau=z,\tau,T]\\*
 &\quad=\frac{1}{p_\tau(z\mid T)}\int k_\tau(z\mid z_0)
       \nabla_{z_0}p_0(z_0\mid T)\dd z_0\\
 &\quad=-\frac{1}{p_\tau(z\mid T)}\int p_0(z_0\mid T)
       \nabla_{z_0}k_\tau(z\mid z_0)\dd z_0\\*
 &\quad=\alpha_\tau\frac{\nabla_z p_\tau(z\mid T)}{p_\tau(z\mid T)}
 =\alpha_\tau s_\tau(z;T).
\end{align*}
\endgroup
Here $\nabla_{z_0}k_\tau=-\alpha_\tau\nabla_z k_\tau$; sufficient decay of $k_\tau p_0$ is assumed for the boundary term to vanish.
To match the residual parameterization~\eqref{eq:residual-score}, define $R_F=-g_F-z_0$, which gives
\begin{equation}
 \begin{aligned}
 \E_{\mathcal P}[R_F\mid z_\tau,\tau,T]
 &=-\alpha_\tau s_\tau(z_\tau;T)
   -\frac{z_\tau+\sigma_\tau^2s_\tau(z_\tau;T)}{\alpha_\tau}\\
 &=-\frac{z_\tau+s_\tau(z_\tau;T)}{\alpha_\tau}.
 \end{aligned}
 \label{eq:force-noisy-target}
\end{equation}
The first equality uses Tweedie's formula~\cite{efron2011}; the final equality uses $\alpha_\tau^2+\sigma_\tau^2=1$.
Thus, $R_F$ has the conditional mean required by Eq.~\eqref{eq:residual-score}, motivating the loss
\begin{equation}
 \mathcal L_F(\theta)=\E_{\mathcal P}\left[\frac{\norm{R_\theta(z_\tau,\tau,T)-R_F}^2}{D}\right].
 \label{eq:force-loss}
\end{equation}

\section{Application to stress calculation at finite temperature}
\label{sec:stress}
To estimate the equilibrium mean Irving--Kirkwood stress in the observation core $C$, we first construct a configuration-based stress estimator.

\subsection{Irving--Kirkwood stress and its spatial average}
Consider a finite-range pair potential that depends only on the distance between atoms; the construction also applies to the effective pair forces of the many-body potential used in Section~\ref{sec:bcc-fe}:
\begin{equation}
 \begin{aligned}
 \widetilde V(q)=\sum_{i<j}\phi(r_{ij}),\qquad
 &r_{ij}=\norm{\bs r_{ij}},\\
 \bs r_{ij}=q_i-q_j,\quad
 &
 \bs f_{ij}=-\phi'(r_{ij})\frac{\bs r_{ij}}{r_{ij}}.
 \end{aligned}
 \label{eq:pair-force}
\end{equation}
Here $\phi(r)$ is the pair energy at separation $r$. The vector $\bs f_{ij}$ is the force on atom $i$ due to atom $j$, so $\bs f_{ji}=-\bs f_{ij}$ and the total force is $\bs f_i=\sum_{j\ne i}\bs f_{ij}$. 

In periodic reference simulations, $\bs r_{ij}$ uses the periodic image associated with the pair interaction.
For generated configurations, it is the direct displacement between the two atoms. With these notations, the instantaneous Irving--Kirkwood stress at spatial position $\bs r$ is~\cite{irving1950,admal2010}
\begin{align}
 \widehat{\bs\sigma}(\bs r)
 ={}&-\sum_i m_i\bs v_i\otimes\bs v_i\,\delta(\bs r-q_i)\nonumber\\
 &-\sum_{i<j}\bs f_{ij}\otimes\bs r_{ij}
           \int_0^1\delta(\bs r-q_j-t\bs r_{ij})\dd t.
 \label{eq:ik-local}
\end{align}
Here $m_i$ is the atomic mass, and $\bs v_i$ is the velocity relative to the macroscopic mean flow. The symbol $\otimes$ denotes the outer product.
In our equilibrium simulations, this mean flow is zero, while atomic velocities fluctuate thermally.
The symbol $\delta$ denotes the Dirac delta distribution, and $t\in[0,1]$ parameterizes the straight bond from $q_j$ to $q_i$.

Our target is the average stress over the core $C$.
Let $|C|$ denote the core area in two dimensions or its volume in three dimensions.
To perform the spatial average, define $\chi_C$ to be one inside $C$ and zero outside, and set
\begin{equation}
 \lambda_{ij,C}=\int_0^1\chi_C(q_j+t\bs r_{ij})\dd t.
 \label{eq:bond-fraction}
\end{equation}
Integrating Eq.~\eqref{eq:ik-local} over $C$ and dividing by $|C|$ then gives
\begin{equation}
 \widehat{\bs\sigma}_C(q,v)
 =-\frac{1}{|C|}\left[
   \sum_i m_i\bs v_i\otimes\bs v_i\chi_C(q_i)
   +\sum_{i<j}\bs f_{ij}\otimes\bs r_{ij}\lambda_{ij,C}\right].
 \label{eq:ik-average}
\end{equation}
Here $v=(\bs v_1,\ldots,\bs v_N)$ collects the atomic velocities, and the hat indicates an instantaneous stress before thermal averaging. In our simulations, the total momentum is unconstrained, and momenta are independent of positions.
The velocities therefore follow the Maxwell--Boltzmann (MB) law, giving
\begin{align*}
 p_{\rm MB}(v\mid T)&\propto\exp\!\left[-\frac\beta2\sum_i m_i\norm{\bs v_i}^2\right],\\
 \E_{v\sim p_{\rm MB}(\cdot\mid T)}[m_i\bs v_i\otimes\bs v_i]&=k_B T\Id.
\end{align*}
Here $\Id=I_d$ is the spatial identity matrix.
Substituting this into Eq.~\eqref{eq:ik-average} gives the estimator used in the experiments:
\begin{equation}
 \overline{\bs\sigma}_C(q,T)
 =-\frac{k_B T N_C(q)}{|C|}\Id
  -\frac{1}{|C|}\sum_{i<j}\bs f_{ij}\otimes\bs r_{ij}\lambda_{ij,C},
 \label{eq:ik-config}
\end{equation}
where $N_C(q)=\sum_i\chi_C(q_i)$ counts the atoms currently inside the core.
The overbar denotes the velocity average at a fixed configuration:
\[
 \overline{\bs\sigma}_C(q,T)=\E_{v\sim p_{\rm MB}(\cdot\mid T)}[\widehat{\bs\sigma}_C(q,v)].
\]
Because atoms can cross the core boundary, $N_C(q)$ can fluctuate and is not replaced by the number of reference lattice sites.
This velocity average preserves the mean kinetic contribution and removes the fluctuations associated with instantaneous velocities.
For aligned configurations, we evaluate the same formula using the positions $q_i-\bar u$, with $C$ fixed in the aligned frame.
Equivalently, the observation window in the original coordinates is $C+\bar u$.
The velocity average is unchanged because alignment affects only positions.

\subsection{Stress equivalence under the marginal effective potential}
With a complete buffer, all atoms in $C$ and both endpoints of each interacting bond intersecting $C$ are retained.
For a many-body potential, the buffer must also contain every neighborhood entering these pair forces.
Then Eq.~\eqref{eq:ik-config} depends only on the aligned retained coordinates $x^{\mathrm a}$.
At fixed $A,T$, we apply the averaging principle of Eq.~\eqref{eq:observable-marginal} to the aligned coordinates.
By the definition of their induced density $p_R^{\mathrm a}$ and its effective potential $V^{\mathrm a}$, the full-system mean stress in the aligned core is
\begin{equation}
 \begin{aligned}
 \langle\bs\sigma_C\rangle_{\Omega}
 &=\E_\Omega[\overline{\bs\sigma}_C(x^{\mathrm a}(q),T)]\\
 &=\int\overline{\bs\sigma}_C(x^{\mathrm a},T)
         p_R^{\mathrm a}(x^{\mathrm a}\mid A,T)\dd x^{\mathrm a}\\
 &=\frac{\displaystyle\int\overline{\bs\sigma}_C(x^{\mathrm a},T)
                         e^{-\beta V^{\mathrm a}(x^{\mathrm a};A,T)}\dd x^{\mathrm a}}
        {\displaystyle\int e^{-\beta V^{\mathrm a}(x^{\mathrm a};A,T)}\dd x^{\mathrm a}}.
 \end{aligned}
 \label{eq:ik-marginal-equivalence}
\end{equation}
Thus, exact marginal sampling preserves the mean core stress.
For $S$ independent samples $x^{\mathrm a,(m)}$ drawn from the generated density $p_{R,\theta}^{\mathrm a}(\cdot\mid T)$, we use
\begin{equation}
 \widehat{\langle\bs\sigma_C\rangle}_{\rm conf}
 =\frac1S\sum_{m=1}^S\overline{\bs\sigma}_C(x^{\mathrm a,(m)},T).
 \label{eq:configuration-route}
\end{equation}
This estimator is unbiased when $p_{R,\theta}^{\mathrm a}=p_R^{\mathrm a}$; in practice, its accuracy depends on the learned distribution, numerical generation, and sample count. All reported GML stress results in the next section use this estimator.

\section{Numerical experiments}
\label{sec:experiments}
We study two crystalline systems containing vacancy clusters. For the two-dimensional Lennard–Jones (LJ) crystal, we examine pair structure, local stresses, training objectives, and sampling performance. For the three-dimensional BCC Fe crystal, we evaluate normal and shear stresses and microscopic mean fields under combined tensile and shear loading. Further implementation details are available at \url{https://github.com/yuancheng-create/GML4MS}.

\subsection{Two-dimensional Lennard--Jones crystal}
\label{sec:lj-demo}
We first use the LJ crystal to assess how accurately GML recovers local equilibrium statistics near a vacancy cluster.

\subsubsection{Numerical setup}
\begin{figure}[!htbp]
 \centering
 \includegraphics[width=\gmlFigureWidth]{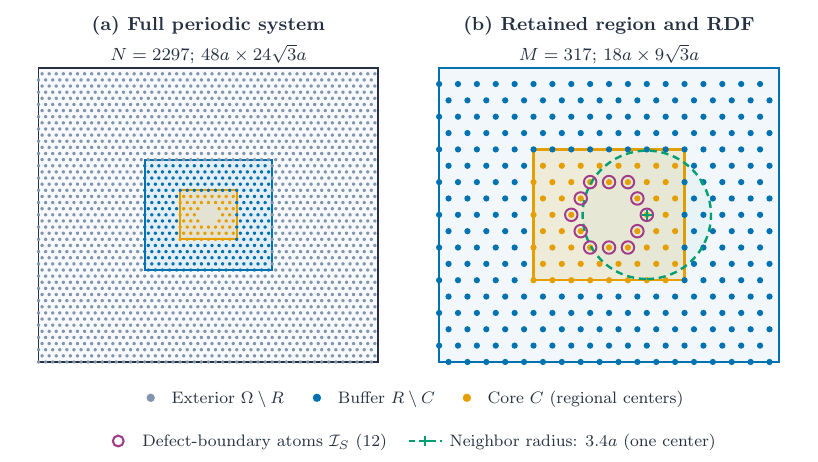}
 \caption{Reference geometry of the seven-vacancy crystal.
 (a) Full periodic system; the blue and orange rectangles delimit $R$ and $C$.
 (b) Retained region and RDF sampling sets.
 Regional centers lie in $C$; purple rings mark the twelve atoms immediately bordering the vacancy cluster, whose labels form $\mathcal I_S$.
 The green plus marks one of these atoms, and its dashed circle shows the neighbor-search radius $3.4a$, where $a$ is the lattice spacing.
 Neighbors may lie in the buffer. The two panels use different scales.}
 \label{fig:computational-system}
\end{figure}

Consider a two-dimensional triangular Lennard--Jones (LJ) crystal with a seven-vacancy cluster, formed by removing the central atom and its six nearest neighbors. Figure~\ref{fig:computational-system} shows the reference geometry and the sampling regions used to compute the radial distribution function (RDF) for validating the pair structure. 
With LJ energy and length scales $\epsilon$ and $\sigma_{\rm LJ}$, respectively, the pair potential is force-shifted at the cutoff radius $r_c$:
\begin{align}
 \phi_{\rm LJ}(r)&=4\epsilon\left[(\sigma_{\rm LJ}/r)^{12}-(\sigma_{\rm LJ}/r)^6\right],\nonumber\\
 \phi(r)&=\begin{cases}
 \phi_{\rm LJ}(r)-\phi_{\rm LJ}(r_c)-(r-r_c)\phi_{\rm LJ}'(r_c),&r<r_c,\\
 0,&r\ge r_c.
 \end{cases}
 \label{eq:lj}
\end{align}
For identical atomic mass $m$, we use reduced units $m=\epsilon=\sigma_{\rm LJ}=k_B=1$ and temperature $T^*=k_BT/\epsilon$. Both energy and force vanish continuously at the cutoff.
Table~\ref{tab:parameters} lists the geometry and numerical parameters.
Here $a$ is the lattice spacing; an $n\times n$ periodic cell has dimensions $na\times n\sqrt3a/2$.
We fix $A=\Id$ and target the marginal of the finite 2297-atom reference system.
All local statistics use the aligned coordinates with observation core
\begin{equation}
 C=[-4a,4a)\times[-2\sqrt3a,2\sqrt3a),\qquad |C|=32\sqrt3a^2.
 \label{eq:core-geometry}
\end{equation}
The nominal buffer widths, $5a$ horizontally and $5\sqrt3a/2$ vertically, exceed the interaction cutoff and the largest RDF radius.

For the MD simulation, we use BAOAB Langevin integration~\cite{leimkuhler2013}.
At each training temperature, six trajectories with continuations supply training data and two independent trajectories supply validation data.
At each evaluation temperature, four large-cell and four small-cell trajectories supply the reference and comparison sets.

\begin{table}[!htbp]
 \centering
 \caption{System and numerical parameters in reduced units. Frame and sample counts are per temperature. MD time is measured in $\sigma_{\rm LJ}\sqrt{m/\epsilon}$. EMA denotes exponential moving average. FP32/FP64 denote 32-/64-bit floating-point formats; BF16 denotes the 16-bit brain floating-point format. AdamW is adaptive moment estimation with decoupled weight decay~\cite{loshchilov2019}.}
 \label{tab:parameters}
 \begingroup\small\setlength{\tabcolsep}{3pt}\renewcommand{\arraystretch}{1.05}
 \begin{tabular*}{\textwidth}{@{\extracolsep{\fill}}*{4}{>{\raggedright\arraybackslash}p{0.235\textwidth}}@{}}
  \toprule
  Parameter & Value & Parameter & Value\\
  \midrule
  \multicolumn{4}{@{}l}{\textbf{Geometry and MD}}\\
  Large cell & $n=48$, $N=2297$ & Retained region & $n=18$, $M=317$\\
  Small periodic cell & $n=8$, 57 atoms & Core reference & 57 atoms\\
  Spacing $a/\sigma_{\rm LJ}$ & $1.11615330299$ & Cutoff $r_c/\sigma_{\rm LJ}$ & $2.5$\\
  MD time step & $0.002$ & Langevin friction & $1$\\
  Equilibration & $10^5$ steps & Production segment & $10^6$ steps\\
  Saving interval & 100 steps & Reference frames & 40000 per MD method\\
  Training $T^*$ & \multicolumn{3}{l}{$0.02,\ 0.04,\ 0.06,\ 0.08,\ 0.10$}\\
  Held-out $T^*$ & \multicolumn{3}{l}{$0.03,\ 0.05,\ 0.07,\ 0.09$}\\
  Training frames & 120000 & Validation frames & 20000\\
  \midrule
  \multicolumn{4}{@{}l}{\textbf{Learning and generation}}\\
  Updates/model & 100000 & Training batch & 256\\
  Optimizer & AdamW & Weight decay & $10^{-5}$\\
  Cosine learning rate & $2\times10^{-4}\!\to\!2\times10^{-5}$ & Gradient-norm clip & 10\\
  EMA decay & $0.999$ & Generated samples & 50000\\
  Heun steps & 25 & Network evaluations & 50\\
  Generation batch & 1000 & Training precision & FP32\\
  Sampling network & BF16 & State/observables & FP32/FP64\\
  \bottomrule
 \end{tabular*}
 \endgroup
\end{table}

We compare three GML models with small periodic MD and a Gaussian baseline, using large-cell MD as the reference. Small periodic MD quantifies finite-size bias, while the Gaussian baseline tests whether GML captures structure and stress beyond the first two moments.
For small periodic MD, $C$ is the whole cell; evaluating $\chi_C$ at periodically wrapped positions gives $\lambda_{ij,C}=1$.

The Gaussian model and GML share the training mean $\mu_T$ and regularized covariance $\Sigma_T=C_T+10^{-9}\operatorname{tr}(C_T)I_D/D$, where $C_T$ is the sample covariance and $D=634$.
Gaussian samples follow $x^{\mathrm a}=\mu_T+L_Tz$, with $z\sim\mathcal N(0,I_D)$ and $L_TL_T^{\mathsf T}=\Sigma_T$.
At held-out temperatures, we linearly interpolate $\mu_T$ and $\Sigma_T/T$ between neighboring training temperatures, then multiply the latter by $T$.
No held-out data enter preprocessing.

The three GML models use force-alignment training ($\mathcal L_F$), denoising training ($\mathcal L_N$), and joint training ($\mathcal L_J=\tfrac12\mathcal L_F+4\mathcal L_N$), respectively.
They share a multiscale convolutional network on the $18\times18$ grid, with nonperiodic padding, reference-position and vacancy-mask inputs, and temperature and noise-level conditioning.
Initial parameters and data/noise draws are matched.
Each objective uses one separately trained network; sampling uses its exponential moving average (EMA) parameters.
Unless stated otherwise, results use the final 100000-update checkpoint. For generation, we integrate the probability-flow ODE~\eqref{eq:learned-probability-flow} using the second-order Heun method~\cite{karras2022elucidating}.

\subsubsection{Numerical results}
\paragraph{Recovery of pair structure}
We first examine RDFs, which test pair structure beyond the Gaussian mean and covariance.

\begin{figure}[!tbp]
 \centering
 \includegraphics[width=\gmlFigureWidth]{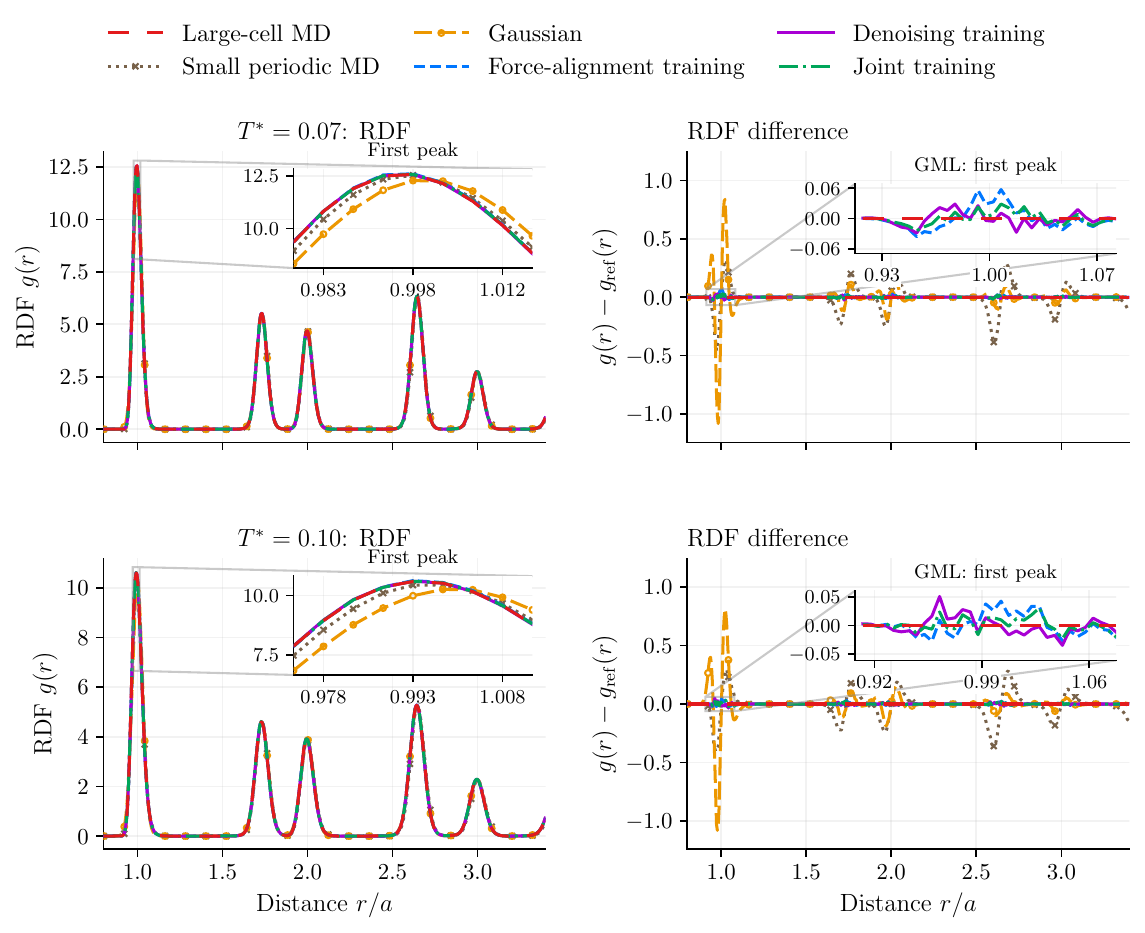}
 \caption{Regional RDFs after 100000 training updates. Rows correspond to $T^*=0.07$ (held out) and $0.10$ (training). The left column shows $g_C$ [Eq.~\eqref{eq:rdf}] and the right column its signed difference from large-cell MD on a linear scale. Insets resolve the first peak and the smaller GML differences.}
 \label{fig:rdf}
\end{figure}
All configurations $x^{\mathrm a,(m)}$ are evaluated in the aligned frame. Define $r_{ij}^{(m)}=\norm{x_i^{\mathrm a,(m)}-x_j^{\mathrm a,(m)}}$. Let bin $b$ have edges $r_b^-,r_b^+$, center $r_b$, and annular area $\Delta A_b=\pi[(r_b^+)^2-(r_b^-)^2]$. With common reference density $\rho_0=2/(\sqrt3a^2)$, the regional RDF is defined as
\begin{equation}
 \widehat g_C(r_b)=
 \frac{\displaystyle\sum_{m=1}^{S}\sum_{i\in\mathcal I_C(x^{\mathrm a,(m)})}
                  \sum_{j\in\mathcal I_R,\,j\ne i}
                  \boldsymbol1\{r_{ij}^{(m)}\in[r_b^-,r_b^+)\}}
      {\displaystyle\rho_0\Delta A_b\sum_{m=1}^{S}N_C(x^{\mathrm a,(m)})},
 \label{eq:rdf}
\end{equation}
where $\mathcal I_C(x^{\mathrm a})=\{i\in\mathcal I_R:x_i^{\mathrm a}\in C\}$, $N_C(x^{\mathrm a})=|\mathcal I_C(x^{\mathrm a})|$. The indicator $\boldsymbol1\{\cdot\}$ equals one when its condition holds and zero otherwise.

The second RDF uses the twelve atoms immediately bordering the seven-vacancy cluster in the reference lattice as centers [Fig.~\ref{fig:computational-system}(b)].
Writing this fixed label set as $\mathcal I_S$, we define
\begin{equation}
 \widehat g_S(r_b)=
 \frac{\displaystyle\sum_{m=1}^{S}\sum_{i\in\mathcal I_S}
                  \sum_{j\in\mathcal I_R,\,j\ne i}
                  \boldsymbol1\{r_{ij}^{(m)}\in[r_b^-,r_b^+)\}}
      {S|\mathcal I_S|\rho_0\Delta A_b}.
 \label{eq:rdf-surface}
\end{equation}
The labels remain fixed during sampling, while the atomic positions fluctuate.
In the figures, $g_C$ and $g_S$ denote the empirical estimators $\widehat g_C$ and $\widehat g_S$.
Both RDFs use bins of width $0.005a$ and a common normalization.
For small periodic MD, $x^{\mathrm a}$ is wrapped into the cell $C$, $\mathcal I_R$ denotes all cell atoms, and distances use minimum images.
For either RDF, the relative discrepancy from the corresponding large-cell estimate is
\begin{equation}
 e_g=
 \left[
 \frac{\displaystyle\sum_{b:\,0.8\le r_b/a\le3.4}
        [\widehat g(r_b)-\widehat g_{\rm ref}(r_b)]^2}
      {\displaystyle\sum_{b:\,0.8\le r_b/a\le3.4}
        \widehat g_{\rm ref}(r_b)^2}
 \right]^{1/2}.
 \label{eq:rdf-error}
\end{equation}

\begin{figure}[!tbp]
 \centering
 \includegraphics[width=\gmlFigureWidth]{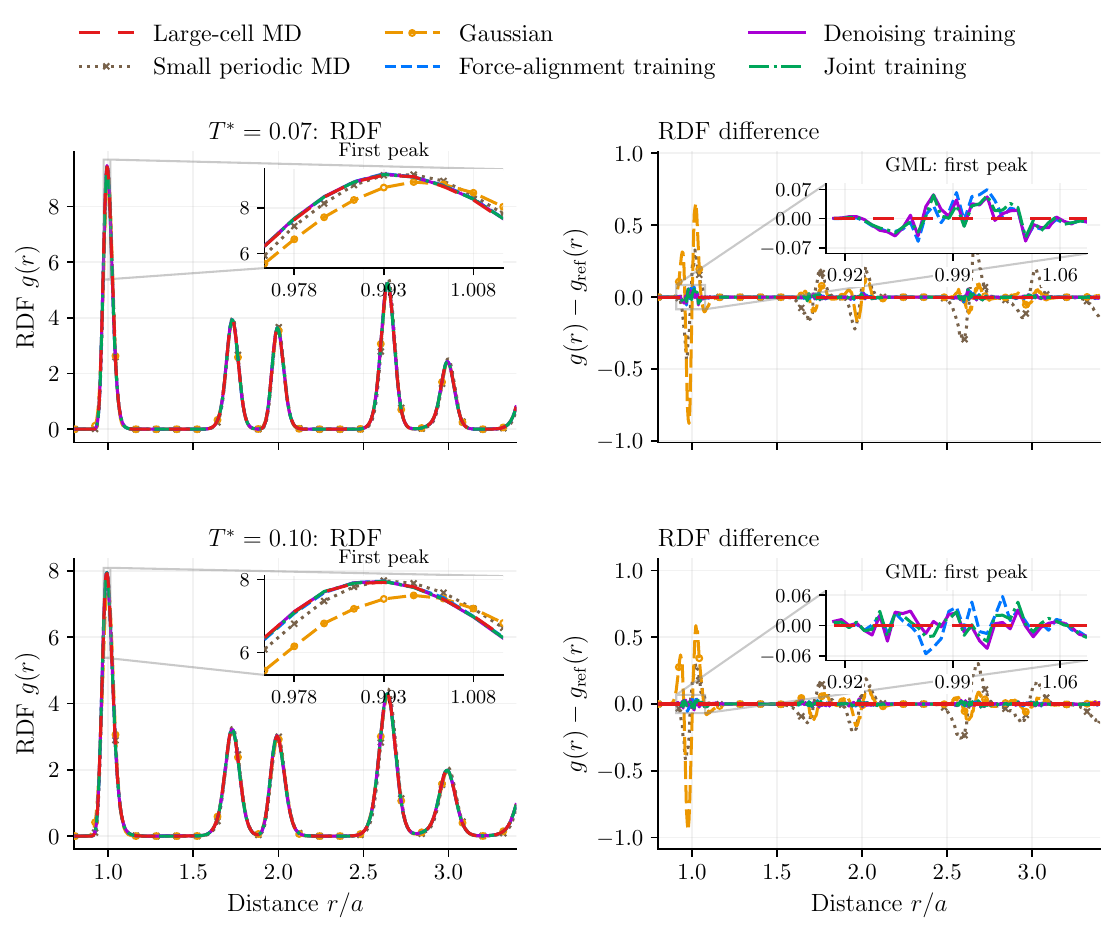}
 \caption{Defect-neighbor RDFs $g_S$ [Eq.~\eqref{eq:rdf-surface}] after 100000 updates, with centers defined by the twelve fixed reference labels. Temperatures, differences, and insets follow Fig.~\ref{fig:rdf}. The centers move with the atoms.}
 \label{fig:rdf-surface}

\medskip
 \begingroup
 \centering
 \captionof{table}{RDF discrepancies over the nine evaluation temperatures after 100000 updates. Entries are the mean $100e_g$ in percent, with the minimum--maximum range in parentheses. The smallest mean in each column is bold.}
 \label{tab:rdf}
 \begingroup\small\setlength{\tabcolsep}{4pt}
 \begin{tabular}{lcc}
  \toprule
  Method & Regional RDF (\%) & Defect-neighbor RDF (\%)\\
  \midrule
  \input{tables/rdf_rows.tex}
  \bottomrule
 \end{tabular}
 \endgroup
 \endgroup
\end{figure}

Figures~\ref{fig:rdf} and~\ref{fig:rdf-surface} show RDFs at a held-out temperature, $T^*=0.07$, and a training temperature, $T^*=0.10$. All methods reproduce the main crystalline peaks, but the difference panels and insets reveal smaller GML errors in both RDFs. Across all nine temperatures (Table~\ref{tab:rdf}), every GML objective improves both RDFs over both baselines. Mean regional discrepancies fall from $4.218\%$ for small periodic MD and $5.334\%$ for the Gaussian model to $0.225$--$0.335\%$ for GML. Defect-neighbor discrepancies fall from $5.053\%$ and $5.631\%$ to $0.457$--$0.501\%$.
Denoising training gives the smallest mean RDF discrepancies, followed closely by joint training.

\begin{figure}[!tbp]
 \centering
 \includegraphics[width=\gmlFigureWidth]{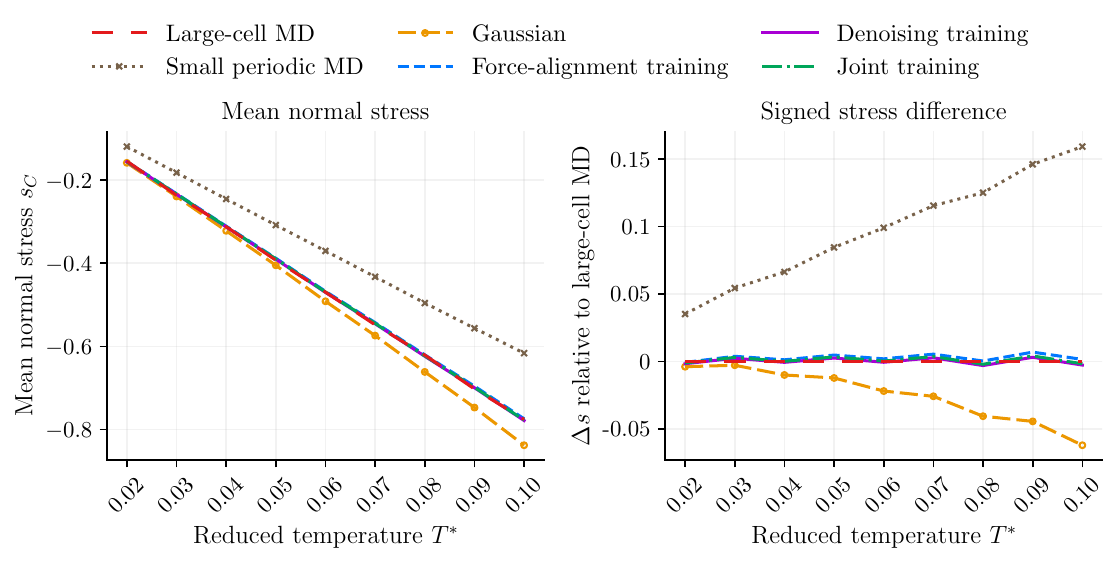}
 \caption{Mean normal stress (left) and signed difference from large-cell MD (right) after 100000 updates. All nine temperatures are shown separately. The right panel retains the sign of the discrepancy, with zero representing agreement with the reference.}
 \label{fig:stress}

\medskip
 \begingroup
 \centering
 \captionof{table}{Reference stress and signed discrepancies after 100000 updates. The reference column retains the original stress units; the five comparison columns report $10^3\Delta s$ [Eq.~\eqref{eq:stress-difference}]. An asterisk marks held-out temperatures. The entry closest to zero in each row is bold.}
 \label{tab:stress}
 \begingroup\small\setlength{\tabcolsep}{4pt}
 \begin{tabular}{rrrrrrr}
  \toprule
  $T^*$ & Reference & Small MD & Gaussian & \shortstack{Force-alignment\\training} & \shortstack{Denoising\\training} & \shortstack{Joint\\training}\\
  \midrule
  \input{tables/stress_rows.tex}
  \bottomrule
 \end{tabular}
 \endgroup
 \endgroup
\end{figure}

\paragraph{Accuracy of local stress}
We now evaluate the mean normal stress and its discrepancy from the reference:
\begin{equation}
 \begin{aligned}
 s_C(x^{\mathrm a},T)&=\tfrac12[\overline\sigma_{C,xx}(x^{\mathrm a},T)+\overline\sigma_{C,yy}(x^{\mathrm a},T)],\\
 \Delta s(T)&=\langle s_C\rangle_{\rm method,T}-\langle s_C\rangle_{\rm ref,T}.
 \end{aligned}
 \label{eq:stress-difference}
\end{equation}
The brackets denote configurational averages for the indicated method at $T$.
GML uses Eq.~\eqref{eq:configuration-route} with the buffered observable~\eqref{eq:ik-config}.
Stress is tensile-positive in units $\epsilon/\sigma_{\rm LJ}^2$; $\Delta s=0$ indicates agreement with the reference.

Figure~\ref{fig:stress} and Table~\ref{tab:stress} show that GML reduces the temperature-dependent biases of both baselines.
Small periodic MD is less compressive than the reference, whereas the Gaussian model is more compressive; both discrepancies grow overall with temperature. Among the three GML models, force-alignment training gives the smallest stress discrepancy at three training temperatures and joint training at two. Denoising training gives the smallest discrepancy at all four held-out temperatures.
This agrees with the RDF evidence for stronger final generalization with denoising training.
However, these final-checkpoint results do not show whether force-alignment training reaches accurate stress estimates faster or whether early stopping can further improve its accuracy.

\begin{figure}[!tbp]
 \centering
 \includegraphics[width=\gmlFigureWidth]{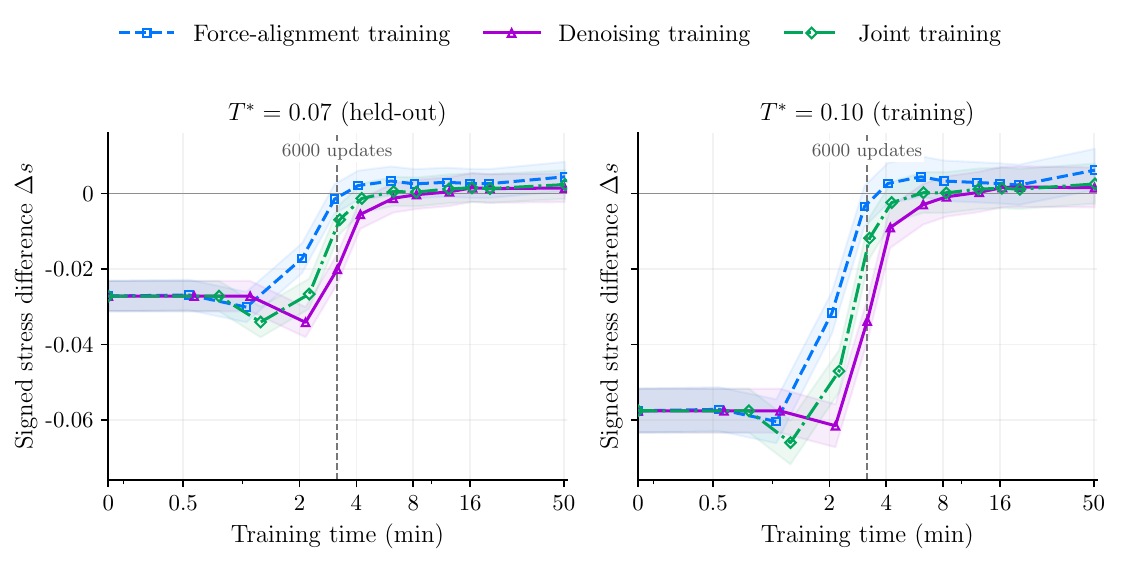}
 \caption{Stress discrepancy during training at $T^*=0.07$ (held out, left) and $0.10$ (training, right). Markers indicate 0, 1000, 2000, 4000, 6000, 8000, 12000, 16000, 24000, 32000, 40000, and 100000 updates. Bands show pointwise 95\% sampling intervals. The time axis is linear below 0.7 minutes and logarithmic thereafter. Dashed vertical lines mark the mean time at 6000 updates.}
 \label{fig:stress-learning}

\end{figure}

\paragraph{Training objectives and early stopping}
\label{sec:training-comparison}
To examine these questions, Fig.~\ref{fig:stress-learning} tracks the stress discrepancy at $T^*=0.07$ and $0.10$ over twelve EMA checkpoints, from initialization to 100000 updates.
Each estimate uses 10000 generated configurations, with shared Gaussian initial vectors and a common 40000-frame MD reference across models and checkpoints at each temperature.
Training time includes validation and checkpoint saving but excludes subsequent generation and stress evaluation.
Pointwise 95\% confidence intervals use 300 bootstrap resamples of blocks of 200 generated configurations and 1000 reference frames.
These intervals describe sampling uncertainty for each fixed network.

The curves show that force-alignment training approaches the reference sooner, then crosses it and drifts away.
By contrast, denoising training converges more steadily, while joint training shows no consistent advantage over either method.
Thus, longer training does not always improve stress accuracy.

To assess the benefit of stopping earlier, we compare the three GML models at 6000 updates, reached after 3.07--3.26 minutes. At this checkpoint, force-alignment training gives the smallest stress discrepancy among the five comparison methods at all nine temperatures (Table~\ref{tab:stress-early}).
Together, these results support force-alignment training for rapid stress estimation, while denoising training gives the best final pair structure and stress generalization in this run.

\begin{table}[!htbp]
 \centering
 \begingroup
 \centering
 \caption{Stress comparison at 6000 updates. Units and temperature notation follow Table~\ref{tab:stress}. Comparison columns give $10^3\Delta s$; bold entries are closest to zero.}
 \label{tab:stress-early}
 \begingroup\small\setlength{\tabcolsep}{4pt}
 \begin{tabular}{rrrrrrr}
  \toprule
  $T^*$ & Reference & Small MD & Gaussian & \shortstack{Force-alignment\\training} & \shortstack{Denoising\\training} & \shortstack{Joint\\training}\\
  \midrule
  0.02 & -0.15437 & +35.144 & -3.946 & \textbf{-3.073} & -6.932 & -4.029 \\
0.03$^{*}$ & -0.23638 & +54.407 & -2.863 & \textbf{+0.722} & -5.417 & -0.969 \\
0.04 & -0.31184 & +66.391 & -10.089 & \textbf{-2.544} & -11.286 & -5.085 \\
0.05$^{*}$ & -0.39268 & +84.584 & -12.297 & \textbf{+0.334} & -11.409 & -3.105 \\
0.06 & -0.46946 & +99.142 & -22.011 & \textbf{-3.246} & -18.212 & -7.662 \\
0.07$^{*}$ & -0.54819 & +115.583 & -25.918 & \textbf{-0.322} & -18.808 & -5.691 \\
0.08 & -0.62075 & +125.153 & -40.674 & \textbf{-6.579} & -28.994 & -12.921 \\
0.09$^{*}$ & -0.70263 & +146.280 & -44.521 & \textbf{-1.188} & -27.358 & -8.408 \\
0.10 & -0.77569 & +159.379 & -62.215 & \textbf{-7.597} & -38.006 & -15.964 \\

  \bottomrule
 \end{tabular}
 \endgroup
 \endgroup

\end{table}

\paragraph{Sampling performance and further acceleration}
\label{sec:sampling-performance}
We now turn from training to sampling cost, using the same geometry for both methods.
MD runs on a central processing unit (CPU), while GML runs on a graphics processing unit (GPU).
Large-cell MD uses twelve single-thread chains, each evolving 2297 atoms for 10000 equilibration steps and 100000 production steps and saving the 317 retained atoms every 100 steps.
Throughput is averaged over three runs of 12000 configurations each, separate from the MD reference data.
The final GML model obtained by denoising training generates 6000 configurations using 25 Heun steps and 50 network evaluations per configuration.

Timings include integration, coordinate restoration, observable evaluation, file output, and the initial recording of network operations for repeated GPU execution. They exclude model loading, compilation, process-pool startup, and offline data acquisition and training. As shown in Table~\ref{tab:performance}, GML achieves a throughput of 2294.72 saved configurations per second, compared with 460.88 for large-cell MD. Because the methods use different hardware and the MD configurations are correlated, these rates should be interpreted as implementation-specific throughput rather than a comparison of the cost per statistically independent sample. On the other hand, further reductions in sampling cost could be achieved by using the learned equilibrium force in Langevin sampling or by applying consistency distillation to reduce the current network evaluations to a single evaluation~\cite{song2023consistency}. We leave these extensions for future work.

\begin{table}[!htbp]
 \centering
 \caption{Saved configurations per second under the conditions described in the text. The larger measured throughput is bold.}
 \label{tab:performance}
 \let\gmlNumericSuperscript\textsuperscript
 \renewcommand{\textsuperscript}[1]{\gmlNumericSuperscript{\ifcase#1\or a\or b\fi}}
 \small
 \begin{tabular}{lr}
  \toprule
  Method & Configurations per second\\
  \midrule
  Large-cell MD\textsuperscript{1} & 460.88 \\
Denoising training\textsuperscript{2} & \textbf{2294.72} \\

  \bottomrule
 \end{tabular}
 \par\smallskip
 \begin{minipage}{\linewidth}\scriptsize\linespread{1}\selectfont\raggedright
 \textsuperscript{1} Intel Core Ultra 9 285K CPU (24 cores); twelve single-thread MD chains.\par
 \textsuperscript{2} One NVIDIA GeForce RTX 5090 D v2 GPU (approximately 24 GB); GML generation batch size 1000.
 \end{minipage}
\end{table}

\FloatBarrier
\subsection{Three-dimensional BCC Fe under combined tension and shear}
\label{sec:bcc-fe}
We then test whether GML extends to a three-dimensional crystal with many-body interactions under combined tension and shear. We compare the training strategies through normal and shear stresses at training and held-out temperatures, then examine atom-resolved configurational stress contributions near the void.

\subsubsection{Numerical setup}
The numerical setup extends the finite-temperature body-centered-cubic (BCC) Fe setting of Dai et al.~\cite{dai2022} to a void under combined tension and shear.
We use the embedded-atom method (EAM) potential of Chamati et al.~\cite{chamati2006}, with the tabulated file provided by Yuri Mishin through the National Institute of Standards and Technology (NIST) Interatomic Potentials Repository.
The zero-temperature lattice constant is $a_0=2.8665$~\AA, with axes along $[100]$, $[010]$, and $[001]$.
The fixed deformation
\begin{equation}
 A=\begin{pmatrix}1.01&0.05&0\\0&1&0\\0&0&1\end{pmatrix},
 \label{eq:bcc-deformation}
\end{equation}
combines $1\%$ tension along $x$ with $5\%$ simple shear.
Removing the 65 sites with $\norm{X_i}\le2a_0$ creates the void.
The full periodic cell, retained region $R$, and core $C$ contain $40^3$, $16^3$, and $6^3$ conventional cells, with 127935, 8127, and 367 reference atoms, respectively.
Regions $R$ and $C$ are the deformed images of $[-8a_0,8a_0)^3$ and $[-3a_0,3a_0)^3$; $|C|=\det(A)(6a_0)^3$ includes the void volume.
Small periodic MD uses the $C$-sized cell with the same defect and deformation.

For the NVT ensemble, we use BAOAB Langevin integration with a 1~fs step, 0.1~ps damping, 50~ps equilibration, and sampling every 0.1~ps.
At each training temperature (50, 150, 250, and 350~K), four training and two independent validation trajectories provide 200~ps of production each.
The test temperatures, 100, 200, 300, and 400~K, are excluded from training, preprocessing, and checkpoint and sampler selection. The first three are interpolation tests; 400~K tests extrapolation beyond the training range.

Motivated by Figure~\ref{fig:stress-learning}, we develop a two-stage training strategy consisting of force alignment followed by denoising (Force-to-denoising) and compare it with force-alignment training (Force) and denoising training (Denoising). All three strategies use the same temperature- and noise-conditioned 3D convolutional architecture on the BCC two-basis grid.
They share initialization, training draws, and an invertible covariance preconditioner.
Training uses AdamW with weight decay $10^{-5}$, batches of 64, a cosine learning-rate schedule from $2\times10^{-4}$ to $2\times10^{-5}$, and EMA decay 0.999.
The Gaussian baseline uses the full empirical covariance with isotropic regularization of $10^{-6}$ times its mean diagonal.
Between training temperatures, we interpolate its mean and its covariance divided by temperature.
At 400~K, the Gaussian baseline uses the 350~K mean and its covariance scaled by 400/350. The network preconditioner likewise fixes its mean at the training endpoint while retaining the square-root temperature scaling of displacements.
All GML results use checkpoints selected by validation-set stress accuracy. Force-to-denoising switches to pure denoising at the validation-selected early force-alignment checkpoint, after 2500 updates.

\subsubsection{Stress accuracy and microscopic mean fields}
\begin{figure}[!htbp]
 \centering
 \includegraphics[width=\gmlFigureWidth]{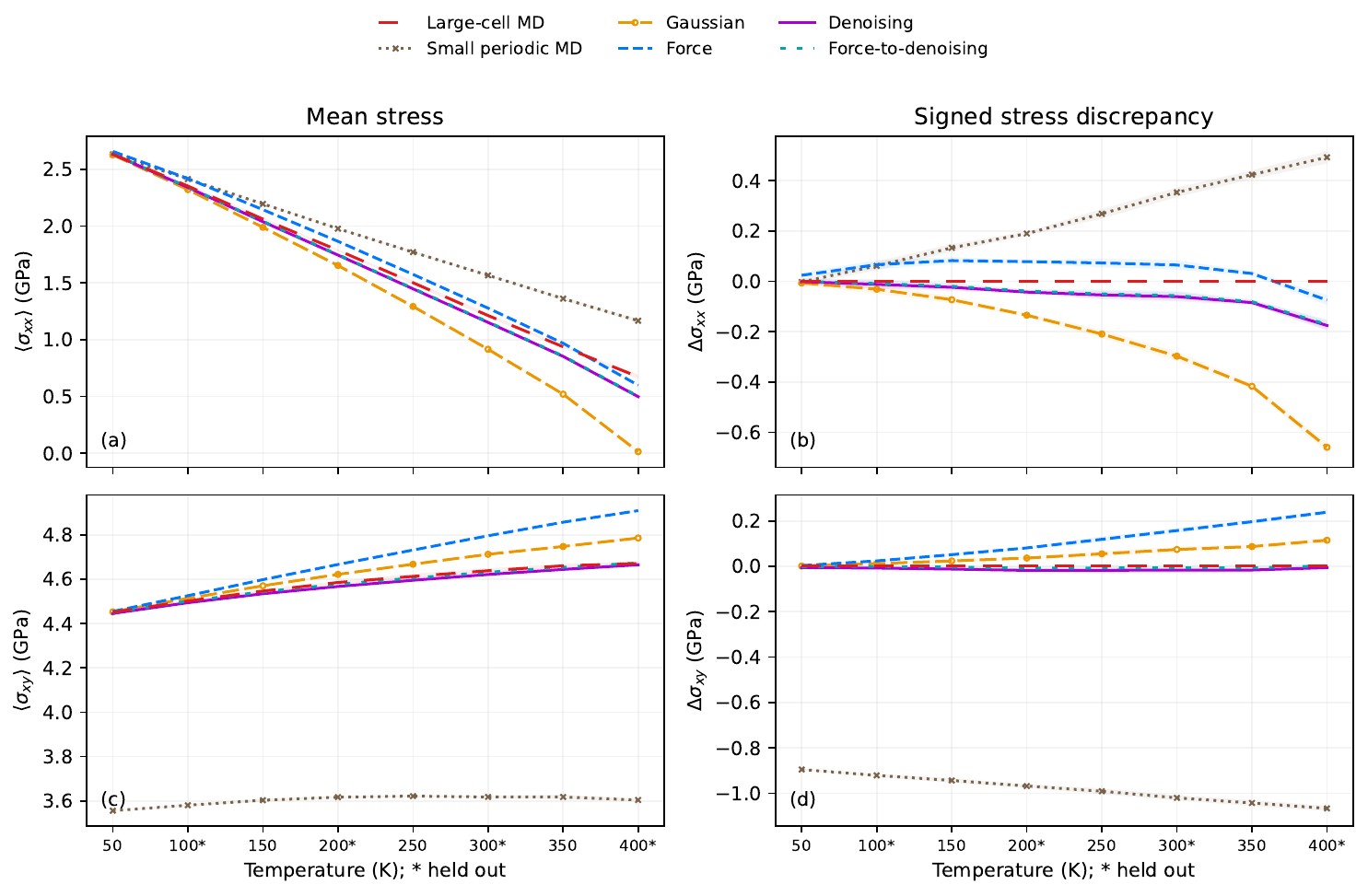}
 \caption{BCC Fe: mean normal ($xx$, top) and shear ($xy$, bottom) stresses (left) and signed discrepancies from large-cell MD (right). Pointwise 95\% interval half-widths for the discrepancies are listed in Tables~\ref{tab:bcc-stress-xx} and~\ref{tab:bcc-stress-xy}. Asterisks mark the test temperatures, excluded from training and model selection; 400~K is an extrapolation test.}
 \label{fig:bcc-stress}
\end{figure}

Figure~\ref{fig:bcc-stress} compares the normal stress $xx$ and shear stress $xy$, with signed discrepancies $\Delta\sigma_{\ell m}=\langle\overline\sigma_{C,\ell m}\rangle_{\rm method}-\langle\overline\sigma_{C,\ell m}\rangle_{\rm ref}$, where $\ell,m\in\{x,y,z\}$ denote Cartesian directions. The components $\overline\sigma_{C,\ell m}$ belong to the velocity-averaged stress in Eq.~\eqref{eq:ik-config}, and the subscript $\mathrm{ref}$ denotes large-cell MD.
When aggregated over the four test temperatures and both stress components, all three GML schedules yield lower root-mean-square (RMS) discrepancies than either baseline (Tables~\ref{tab:bcc-stress-xx} and~\ref{tab:bcc-stress-xy}).
Meanwhile, the improvement is not uniform across objectives: force alignment gives the smallest normal-stress discrepancy at 350 and 400~K, but its shear discrepancy grows with temperature. Denoising and force-to-denoising keep the absolute shear discrepancy below 0.02~GPa at all eight temperatures.

\begin{table}[!htbp]
 \centering
 \caption{BCC Fe: signed $xx$ stress discrepancies (GPa). The $\pm$ values are pointwise 95\% interval half-widths; asterisks mark test temperatures. Bold entries have the smallest absolute discrepancy at the displayed precision; this ranking does not establish statistical significance.}
 \label{tab:bcc-stress-xx}
 \begingroup\scriptsize\setlength{\tabcolsep}{2.5pt}\renewcommand{\arraystretch}{1.2}
 \begin{tabular}{rccccc}
 \toprule
 $T$ (K) & Force & Denoising & \shortstack{Force-to-\\denoising} & Gaussian & \shortstack{Small\\periodic MD}\\
 \midrule
 50 & $+0.0241\pm0.0111$ & $-0.0011\pm0.0111$ & $\mathbf{+0.0003}\pm0.0111$ & $-0.0064\pm0.0111$ & $-0.0023\pm0.0111$ \\
100$^{*}$ & $+0.0664\pm0.0175$ & $-0.0118\pm0.0176$ & $\mathbf{-0.0086}\pm0.0176$ & $-0.0311\pm0.0176$ & $+0.0617\pm0.0176$ \\
150 & $+0.0827\pm0.0140$ & $-0.0235\pm0.0141$ & $\mathbf{-0.0193}\pm0.0141$ & $-0.0729\pm0.0141$ & $+0.1334\pm0.0149$ \\
200$^{*}$ & $+0.0784\pm0.0073$ & $-0.0426\pm0.0076$ & $\mathbf{-0.0382}\pm0.0076$ & $-0.1344\pm0.0077$ & $+0.1901\pm0.0080$ \\
250 & $+0.0734\pm0.0170$ & $-0.0539\pm0.0171$ & $\mathbf{-0.0498}\pm0.0171$ & $-0.2092\pm0.0172$ & $+0.2688\pm0.0191$ \\
300$^{*}$ & $+0.0650\pm0.0183$ & $-0.0606\pm0.0185$ & $\mathbf{-0.0563}\pm0.0185$ & $-0.2973\pm0.0186$ & $+0.3539\pm0.0188$ \\
350 & $\mathbf{+0.0312}\pm0.0096$ & $-0.0844\pm0.0099$ & $-0.0806\pm0.0099$ & $-0.4170\pm0.0103$ & $+0.4242\pm0.0171$ \\
400$^{*}$ & $\mathbf{-0.0745}\pm0.0198$ & $-0.1754\pm0.0199$ & $-0.1715\pm0.0199$ & $-0.6593\pm0.0203$ & $+0.4933\pm0.0241$ \\

 \bottomrule
 \end{tabular}
 \endgroup
\end{table}

\begin{table}[!htbp]
 \centering
 \caption{BCC Fe: signed $xy$ stress discrepancies (GPa). The $\pm$ values are pointwise 95\% interval half-widths; asterisks mark test temperatures. Bold entries have the smallest absolute discrepancy at the displayed precision; this ranking does not establish statistical significance.}
 \label{tab:bcc-stress-xy}
 \begingroup\scriptsize\setlength{\tabcolsep}{2.5pt}\renewcommand{\arraystretch}{1.2}
 \begin{tabular}{rccccc}
 \toprule
 $T$ (K) & Force & Denoising & \shortstack{Force-to-\\denoising} & Gaussian & \shortstack{Small\\periodic MD}\\
 \midrule
 50 & $+0.0032\pm0.0034$ & $-0.0058\pm0.0034$ & $-0.0014\pm0.0035$ & $\mathbf{+0.0013}\pm0.0034$ & $-0.8956\pm0.0035$ \\
100$^{*}$ & $+0.0239\pm0.0053$ & $-0.0079\pm0.0054$ & $\mathbf{-0.0007}\pm0.0054$ & $+0.0118\pm0.0054$ & $-0.9214\pm0.0069$ \\
150 & $+0.0513\pm0.0085$ & $-0.0123\pm0.0086$ & $\mathbf{-0.0035}\pm0.0086$ & $+0.0237\pm0.0086$ & $-0.9437\pm0.0106$ \\
200$^{*}$ & $+0.0810\pm0.0056$ & $-0.0181\pm0.0058$ & $\mathbf{-0.0081}\pm0.0058$ & $+0.0364\pm0.0058$ & $-0.9684\pm0.0068$ \\
250 & $+0.1189\pm0.0126$ & $-0.0175\pm0.0127$ & $\mathbf{-0.0072}\pm0.0127$ & $+0.0550\pm0.0128$ & $-0.9912\pm0.0129$ \\
300$^{*}$ & $+0.1575\pm0.0111$ & $-0.0170\pm0.0113$ & $\mathbf{-0.0068}\pm0.0113$ & $+0.0741\pm0.0114$ & $-1.0204\pm0.0116$ \\
350 & $+0.1965\pm0.0091$ & $-0.0165\pm0.0093$ & $\mathbf{-0.0069}\pm0.0093$ & $+0.0870\pm0.0096$ & $-1.0429\pm0.0128$ \\
400$^{*}$ & $+0.2384\pm0.0103$ & $-0.0058\pm0.0105$ & $\mathbf{+0.0023}\pm0.0105$ & $+0.1150\pm0.0109$ & $-1.0669\pm0.0114$ \\

 \bottomrule
 \end{tabular}
 \endgroup
\end{table}

\begin{samepage}
As shown in Tables~\ref{tab:bcc-stress-xx} and~\ref{tab:bcc-stress-xy}, Force-to-denoising yields the smallest absolute stress discrepancies for both components at all three interpolation temperatures. At the extrapolation temperature of 400~K, Force gives the most accurate normal stress, whereas Force-to-denoising achieves the smallest shear-stress error and the lowest combined root-mean-square (RMS) error. These results support the ability of our method to generate samples at a temperature beyond the training range.
\par
\end{samepage}

\begin{figure}[!htbp]
 \centering
 \includegraphics[width=\gmlFigureWidth]{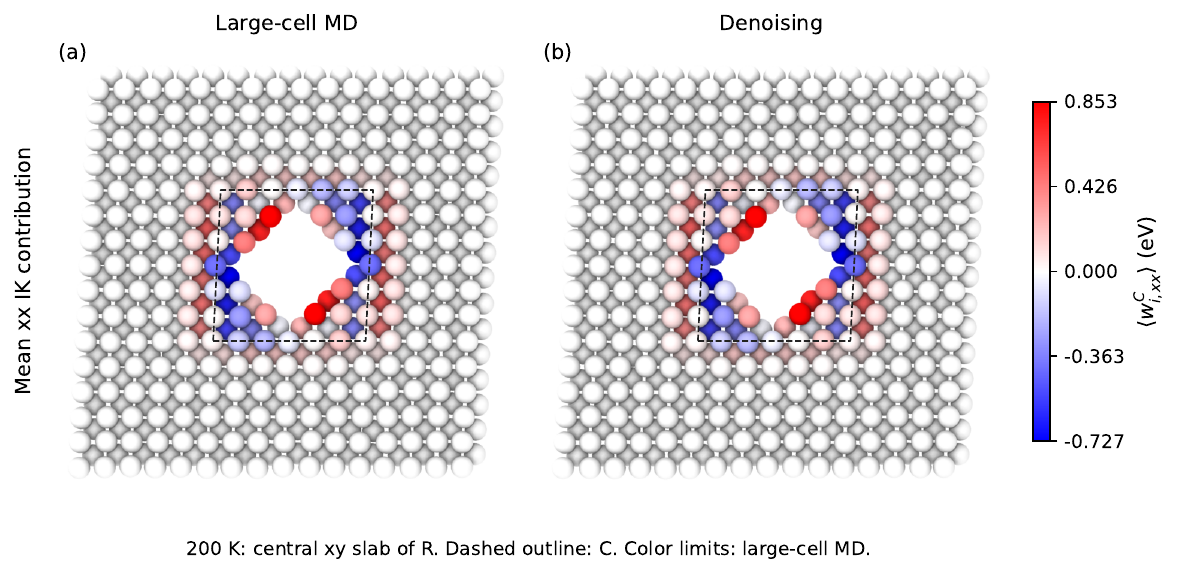}
 \caption{Mean $xx$ IK contributions $\langle w_{i,xx}^{C}\rangle$ to stress in $C$ at 200~K [Eq.~\eqref{eq:bcc-atomic-contribution}]: large-cell MD (left) and denoising GML (right). Each panel shows one instantaneous configuration colored by per-atom ensemble means. The view contains 487 labels whose reference positions satisfy $-a_0/2\le X_{i,z}<a_0/2$ in the central slab of $R$; dashed lines delimit $C$. Both panels use the same color scale set by the MD extrema over $R$, with zero shown in white.}
 \label{fig:bcc-local-fields}
\end{figure}

Agreement in core stress could still mask compensating atomic errors. We therefore compare atom-resolved mean configurational stress contributions from large-cell MD and denoising GML at 200~K (Fig.~\ref{fig:bcc-local-fields}).

\begin{samepage}
The per-atom IK contribution is
\begin{equation}
 \begin{aligned}
 w_{i,xx}^{C}(q)&=-\frac12\sum_{j\ne i}
       \lambda_{ij,C}(q)f_{ij,x}(q)r_{ij,x}(q),\\
 \sigma^{\rm conf}_{C,xx}(q)&=\frac1{|C|}\sum_{i\in\mathcal I_R}w_{i,xx}^{C}(q).
 \end{aligned}
 \label{eq:bcc-atomic-contribution}
\end{equation}
Here $q$ is the configuration of Section~\ref{sec:marginal} [Eq.~\eqref{eq:displacement}], expressed in the aligned frame of Eq.~\eqref{eq:aligned-coordinates}.
\par
\end{samepage}

With a complete buffer, only the retained coordinates $x^{\mathrm a}$ are needed. The bond fraction $\lambda_{ij,C}$ is defined in Eq.~\eqref{eq:bond-fraction}; $f_{ij,x}$ and $r_{ij,x}$ are the $x$ components of the effective EAM pair force and bond displacement, respectively. The superscript $\mathrm{conf}$ denotes the configurational part of Eq.~\eqref{eq:ik-config}.
Bond contributions are computed before ensemble averaging and shared equally by their endpoints, including those outside $C$.
Thus $w_{i,xx}^{C}$, in eV, resolves the configurational stress in $C$ over atoms in $R$ and excludes the kinetic term.

The colors represent $\langle w_{i,xx}^{C}\rangle$, averaged over 16000 MD and 20000 generated configurations.
Denoising GML captures the tensile and compressive contributions around the void.
For the 367 core labels, the root-mean-square discrepancy is $0.00416$~eV, with a maximum absolute discrepancy of $0.0106$~eV. This atom-resolved comparison supports the recovery of local stress contributions beyond agreement in core-averaged stress.

\section{Conclusions}
\label{sec:conclusions}
We developed GML to learn the equilibrium marginal distribution of a retained atomic region from large-system simulations using denoising and force-alignment training. The theoretical formulation establishes that exact sampling from this marginal preserves the equilibrium mean Irving--Kirkwood (IK) stress in the observation core. The LJ experiments demonstrate improved pair structure and mean stress relative to small-cell periodic MD and a Gaussian baseline. The BCC Fe example extends the stress calculation to three dimensions and mixed deformation. In the LJ study, the two training strategies exhibit complementary strengths: force alignment achieves accurate stress estimates earlier in training, whereas denoising yields better generalization at convergence. Under the hardware and timing conditions reported for the LJ study, GML also achieves higher configuration-generation throughput than large-cell MD, although these rates do not establish a lower cost per statistically independent sample. These results support retained-region marginal learning as an effective approach to incorporating the statistical influence of the surrounding material while preserving an explicit atomistic description within the region of interest. Future work will focus on accurate score estimation in the zero-noise limit, more efficient sampling, and extensions to a broader range of deformation states, defect types, and multiscale modeling frameworks.

\section*{Acknowledgments}
This work was supported by the National Natural Science Foundation of China (No.~12301558).

\bibliographystyle{elsarticle-num}
\bibliography{references}

@inproceedings{liu2024marginalization,
  author = {Liu, Sulin and Ramadge, Peter J. and Adams, Ryan P.},
  title = {Generative marginalization models},
  booktitle = {Proceedings of the 41st International Conference on Machine Learning},
  series = {Proceedings of Machine Learning Research},
  volume = {235},
  pages = {31773--31807},
  publisher = {PMLR},
  year = {2024},
  url = {https://proceedings.mlr.press/v235/liu24az.html}
}

@article{thiemann2026,
  author = {Thiemann, Fabian L. and Resch{\"u}tzegger, Thiago and Esposito, Massimiliano and Taddese, Tseden and Olarte-Plata, Juan D. and Martelli, Fausto},
  title = {Force-free molecular dynamics through autoregressive equivariant networks},
  journal = {Nat. Mach. Intell.},
  volume = {8},
  pages = {764--776},
  year = {2026},
  doi = {10.1038/s42256-026-01227-7}
}

@misc{mccallum2026,
  author = {McCallum, Sam and Rindtorff, Niklas and Tong, Alexander and Foster, James},
  title = {{Langevin} flow maps: Efficient molecular dynamics and transition path sampling},
  howpublished = {arXiv preprint arXiv:2610.05998},
  year = {2026},
  url = {https://arxiv.org/abs/2610.05998}
}

@misc{xie2026,
  author = {Xie, Yu and Winkler, Ludwig and Sun, Lixin and Lewis, Sarah and Foster, Adam E. and {Jim{\'e}nez Luna}, Jos{\'e} and Hempel, Tim and Gastegger, Michael and Chen, Yaoyi and Zaporozhets, Iryna and Clementi, Cecilia and Bishop, Christopher M. and No{\'e}, Frank},
  title = {Enhanced diffusion sampling: Efficient rare event sampling and free energy calculation with diffusion models},
  howpublished = {arXiv preprint arXiv:2602.16634},
  year = {2026},
  url = {https://arxiv.org/abs/2602.16634}
}

@book{frenkel2002,
  author = {Frenkel, Daan and Smit, Berend},
  title = {Understanding Molecular Simulation: From Algorithms to Applications},
  edition = {2nd},
  publisher = {Academic Press},
  address = {San Diego},
  year = {2002},
  isbn = {9780122673511},
  doi = {10.1016/B978-0-12-267351-1.X5000-7}
}

@article{abraham1994,
  author = {Abraham, Farid F. and Brodbeck, D. and Rafey, R. A. and Rudge, W. E.},
  title = {Instability dynamics of fracture: A computer simulation investigation},
  journal = {Phys. Rev. Lett.},
  volume = {73},
  number = {2},
  pages = {272--275},
  year = {1994},
  doi = {10.1103/PhysRevLett.73.272}
}

@article{schiotz1998,
  author = {Schi{\o}tz, Jakob and {Di Tolla}, Francesco D. and Jacobsen, Karsten W.},
  title = {Softening of nanocrystalline metals at very small grain sizes},
  journal = {Nature},
  volume = {391},
  pages = {561--563},
  year = {1998},
  doi = {10.1038/35328}
}

@article{noid2008,
  author = {Noid, W. G. and Chu, Jhih-Wei and Ayton, Gary S. and Krishna, Vinod and Izvekov, Sergei and Voth, Gregory A. and Das, Avisek and Andersen, Hans C.},
  title = {The multiscale coarse-graining method. {I}. A rigorous bridge between atomistic and coarse-grained models},
  journal = {J. Chem. Phys.},
  volume = {128},
  pages = {244114},
  year = {2008},
  doi = {10.1063/1.2938860}
}

@article{vincent2011,
  author = {Vincent, Pascal},
  title = {A connection between score matching and denoising autoencoders},
  journal = {Neural Comput.},
  volume = {23},
  pages = {1661--1674},
  year = {2011},
  doi = {10.1162/NECO_a_00142}
}

@misc{debortoli2024,
  author = {{De Bortoli}, Valentin and Hutchinson, Michael and Wirnsberger, Peter and Doucet, Arnaud},
  title = {Target score matching},
  howpublished = {arXiv preprint arXiv:2402.08667},
  year = {2024},
  url = {https://arxiv.org/abs/2402.08667}
}

@misc{ko2026,
  author = {Ko, Joohwan and Geffner, Tomas},
  title = {Latent target score matching, with an application to simulation-based inference},
  howpublished = {arXiv preprint arXiv:2602.07189},
  year = {2026},
  url = {https://arxiv.org/abs/2602.07189}
}

@article{efron2011,
  author = {Efron, Bradley},
  title = {{Tweedie}'s formula and selection bias},
  journal = {J. Am. Stat. Assoc.},
  volume = {106},
  number = {496},
  pages = {1602--1614},
  year = {2011},
  doi = {10.1198/jasa.2011.tm11181}
}

@inproceedings{ho2020,
  author = {Ho, Jonathan and Jain, Ajay and Abbeel, Pieter},
  title = {Denoising diffusion probabilistic models},
  booktitle = {Advances in Neural Information Processing Systems},
  volume = {33},
  pages = {6840--6851},
  publisher = {Curran Associates, Inc.},
  year = {2020},
  url = {https://arxiv.org/abs/2006.11239}
}

@misc{song2021,
  author = {Song, Yang and Sohl-Dickstein, Jascha and Kingma, Diederik P. and Kumar, Abhishek and Ermon, Stefano and Poole, Ben},
  title = {Score-based generative modeling through stochastic differential equations},
  howpublished = {International Conference on Learning Representations},
  year = {2021},
  url = {https://openreview.net/forum?id=PxTIG12RRHS},
  eprint = {2011.13456},
  archivePrefix = {arXiv}
}

@inproceedings{song2023consistency,
  author = {Song, Yang and Dhariwal, Prafulla and Chen, Mark and Sutskever, Ilya},
  title = {Consistency models},
  booktitle = {Proceedings of the 40th International Conference on Machine Learning},
  series = {Proceedings of Machine Learning Research},
  volume = {202},
  pages = {32211--32252},
  publisher = {PMLR},
  year = {2023},
  url = {https://proceedings.mlr.press/v202/song23a.html}
}

@inproceedings{nichol2021,
  author = {Nichol, Alexander Quinn and Dhariwal, Prafulla},
  title = {Improved denoising diffusion probabilistic models},
  booktitle = {Proceedings of the 38th International Conference on Machine Learning},
  series = {Proceedings of Machine Learning Research},
  volume = {139},
  pages = {8162--8171},
  publisher = {PMLR},
  year = {2021},
  url = {https://proceedings.mlr.press/v139/nichol21a.html}
}

@inproceedings{vahdat2021,
  author = {Vahdat, Arash and Kreis, Karsten and Kautz, Jan},
  title = {Score-based generative modeling in latent space},
  booktitle = {Advances in Neural Information Processing Systems},
  volume = {34},
  pages = {11287--11302},
  publisher = {Curran Associates},
  year = {2021},
  url = {https://arxiv.org/abs/2106.05931}
}

@article{irving1950,
  author = {Irving, J. H. and Kirkwood, John G.},
  title = {The statistical mechanical theory of transport processes. {IV}. The equations of hydrodynamics},
  journal = {J. Chem. Phys.},
  volume = {18},
  pages = {817--829},
  year = {1950},
  doi = {10.1063/1.1747782}
}

@article{admal2010,
  author = {Admal, Nikhil Chandra and Tadmor, Ellad B.},
  title = {A unified interpretation of stress in molecular systems},
  journal = {J. Elasticity},
  volume = {100},
  pages = {63--143},
  year = {2010},
  doi = {10.1007/s10659-010-9249-6}
}

@article{tadmor1996,
  author = {Tadmor, E. B. and Ortiz, M. and Phillips, R.},
  title = {Quasicontinuum analysis of defects in solids},
  journal = {Philos. Mag. A},
  volume = {73},
  number = {6},
  pages = {1529--1563},
  year = {1996},
  doi = {10.1080/01418619608243000}
}

@article{e2003,
  author = {E, Weinan and Engquist, Bj{\"o}rn and Huang, Zhongyi},
  title = {Heterogeneous multiscale method: A general methodology for multiscale modeling},
  journal = {Phys. Rev. B},
  volume = {67},
  pages = {092101},
  year = {2003},
  doi = {10.1103/PhysRevB.67.092101}
}

@article{cai2003,
  author = {Cai, Wei and Bulatov, Vasily V. and Chang, Jinpeng and Li, Ju and Yip, Sidney},
  title = {Periodic image effects in dislocation modelling},
  journal = {Philos. Mag.},
  volume = {83},
  number = {5},
  pages = {539--567},
  year = {2003},
  doi = {10.1080/0141861021000051109}
}

@article{li2006,
  author = {Li, Xiantao and E, Weinan},
  title = {Variational boundary conditions for molecular dynamics simulations of solids at low temperature},
  journal = {Commun. Comput. Phys.},
  volume = {1},
  number = {1},
  pages = {135--175},
  year = {2006},
  doi = {10.4208/cicp.2006.v1.p135}
}

@article{yang2006,
  author = {Yang, Jerry Z. and Li, Xiantao},
  title = {Comparative study of boundary conditions for molecular dynamics simulations of solids at low temperature},
  journal = {Phys. Rev. B},
  volume = {73},
  pages = {224111},
  year = {2006},
  doi = {10.1103/PhysRevB.73.224111}
}

@article{yuan2019,
  author = {Yuan, Cheng and Yang, Jerry Zhijian and Li, Xiantao},
  title = {An effective and easy-to-implement boundary condition for molecular dynamics simulations},
  journal = {Commun. Comput. Phys.},
  volume = {26},
  number = {1},
  pages = {192--205},
  year = {2019},
  doi = {10.4208/cicp.OA-2018-0303}
}

@article{li2007,
  author = {Li, Xiantao and E, Weinan},
  title = {Variational boundary conditions for molecular dynamics simulations of crystalline solids at finite temperature: Treatment of the thermal bath},
  journal = {Phys. Rev. B},
  volume = {76},
  pages = {104107},
  year = {2007},
  doi = {10.1103/PhysRevB.76.104107}
}

@article{dupuy2005,
  author = {Dupuy, L. M. and Tadmor, E. B. and Miller, R. E. and Phillips, R.},
  title = {Finite-temperature quasicontinuum: Molecular dynamics without all the atoms},
  journal = {Phys. Rev. Lett.},
  volume = {95},
  pages = {060202},
  year = {2005},
  doi = {10.1103/PhysRevLett.95.060202}
}

@article{behler2007,
  author = {Behler, J{\"o}rg and Parrinello, Michele},
  title = {Generalized neural-network representation of high-dimensional potential-energy surfaces},
  journal = {Phys. Rev. Lett.},
  volume = {98},
  pages = {146401},
  year = {2007},
  doi = {10.1103/PhysRevLett.98.146401}
}

@article{zhang2018,
  author = {Zhang, Linfeng and Han, Jiequn and Wang, Han and Car, Roberto and E, Weinan},
  title = {{Deep Potential Molecular Dynamics}: A scalable model with the accuracy of quantum mechanics},
  journal = {Phys. Rev. Lett.},
  volume = {120},
  pages = {143001},
  year = {2018},
  doi = {10.1103/PhysRevLett.120.143001}
}

@article{leimkuhler2013,
  author = {Leimkuhler, Benedict and Matthews, Charles},
  title = {Robust and efficient configurational molecular sampling via {Langevin} dynamics},
  journal = {J. Chem. Phys.},
  volume = {138},
  pages = {174102},
  year = {2013},
  doi = {10.1063/1.4802990}
}

@misc{loshchilov2019,
  author = {Loshchilov, Ilya and Hutter, Frank},
  title = {Decoupled weight decay regularization},
  howpublished = {International Conference on Learning Representations},
  year = {2019},
  url = {https://arxiv.org/abs/1711.05101}
}

@article{batzner2022,
  author = {Batzner, Simon and Musaelian, Albert and Sun, Lixin and Geiger, Mario and Mailoa, Jonathan P. and Kornbluth, Mordechai and Molinari, Nicola and Smidt, Tess E. and Kozinsky, Boris},
  title = {{E(3)}-equivariant graph neural networks for data-efficient and accurate interatomic potentials},
  journal = {Nat. Commun.},
  volume = {13},
  pages = {2453},
  year = {2022},
  doi = {10.1038/s41467-022-29939-5}
}

@inproceedings{batatia2022,
  author = {Batatia, Ilyes and Kov{\'a}cs, D{\'a}vid P{\'e}ter and Simm, Gregor N. C. and Ortner, Christoph and Cs{\'a}nyi, G{\'a}bor},
  title = {{MACE}: Higher order equivariant message passing neural networks for fast and accurate force fields},
  booktitle = {Advances in Neural Information Processing Systems},
  volume = {35},
  pages = {11423--11436},
  year = {2022},
  doi = {10.52202/068431-0830}
}

@article{ji2025,
  author = {Ji, Yajie and Liang, Jiuyang and Xu, Zhenli},
  title = {Machine-learning interatomic potentials for long-range systems},
  journal = {Phys. Rev. Lett.},
  volume = {135},
  pages = {178001},
  year = {2025},
  doi = {10.1103/ssp9-7s81}
}

@article{kessy2018,
  author = {Kessy, Agnan and Lewin, Alex and Strimmer, Korbinian},
  title = {Optimal whitening and decorrelation},
  journal = {Am. Stat.},
  volume = {72},
  number = {4},
  pages = {309--314},
  year = {2018},
  doi = {10.1080/00031305.2016.1277159}
}

@article{noe2019,
  author = {No{\'e}, Frank and Olsson, Simon and K{\"o}hler, Jonas and Wu, Hao},
  title = {Boltzmann generators: Sampling equilibrium states of many-body systems with deep learning},
  journal = {Science},
  volume = {365},
  pages = {eaaw1147},
  year = {2019},
  doi = {10.1126/science.aaw1147}
}

@article{ahmad2022,
  author = {Ahmad, Rasool and Cai, Wei},
  title = {Free energy calculation of crystalline solids using normalizing flows},
  journal = {Modell. Simul. Mater. Sci. Eng.},
  volume = {30},
  pages = {065007},
  year = {2022},
  doi = {10.1088/1361-651X/ac7f4b}
}

@article{kohler2023,
  author = {K{\"o}hler, Jonas and Chen, Yaoyi and Kr{\"a}mer, Andreas and Clementi, Cecilia and No{\'e}, Frank},
  title = {Flow-matching: Efficient coarse-graining of molecular dynamics without forces},
  journal = {J. Chem. Theory Comput.},
  volume = {19},
  number = {3},
  pages = {942--952},
  year = {2023},
  doi = {10.1021/acs.jctc.3c00016}
}

@article{dai2022,
  author = {Dai, Shuyang and Wang, Fengru and Yang, Jerry Zhijian and Yuan, Cheng},
  title = {On the {Cauchy--Born} approximation at finite temperature for alloys},
  journal = {Discrete Contin. Dyn. Syst. B},
  volume = {27},
  number = {6},
  pages = {3131--3153},
  year = {2022},
  doi = {10.3934/dcdsb.2021176}
}

@article{arts2023,
  author = {Arts, Marloes and Garc{\'i}a Satorras, Victor and Huang, Chin-Wei and Z{\"u}gner, Daniel and Federici, Marco and Clementi, Cecilia and No{\'e}, Frank and Pinsler, Robert and van den Berg, Rianne},
  title = {Two for One: Diffusion models and force fields for coarse-grained molecular dynamics},
  journal = {J. Chem. Theory Comput.},
  volume = {19},
  number = {18},
  pages = {6151--6159},
  year = {2023},
  doi = {10.1021/acs.jctc.3c00702}
}

@article{kalligiannaki2015,
  author = {Kalligiannaki, Evangelia and Harmandaris, Vagelis and Katsoulakis, Markos A. and Plech{\'a}{\v{c}}, Petr},
  title = {The geometry of generalized force matching and related information metrics in coarse-graining of molecular systems},
  journal = {J. Chem. Phys.},
  volume = {143},
  number = {8},
  pages = {084105},
  year = {2015},
  doi = {10.1063/1.4928857}
}

@inproceedings{karras2022elucidating,
  author = {Karras, Tero and Aittala, Miika and Aila, Timo and Laine, Samuli},
  title = {Elucidating the Design Space of Diffusion-Based Generative Models},
  booktitle = {Advances in Neural Information Processing Systems},
  volume = {35},
  pages = {26565--26577},
  publisher = {Curran Associates, Inc.},
  year = {2022},
  doi = {10.52202/068431-1926}
}

@article{chamati2006,
  author = {Chamati, H. and Papanicolaou, N. I. and Mishin, Y. and Papaconstantopoulos, D. A.},
  title = {Embedded-atom potential for {Fe} and its application to self-diffusion on {Fe}(100)},
  journal = {Surface Science},
  volume = {600},
  number = {9},
  pages = {1793--1803},
  year = {2006},
  doi = {10.1016/j.susc.2006.02.010}
}
\end{document}